\documentclass[11pt]{article}
\usepackage{graphicx}
\usepackage{amsmath}
\usepackage{amssymb}
\usepackage{amsxtra}

\begin{document}

\begin{flushright} 
OIQP-14-10
\end{flushright} 

\begin{Large}
\vspace{1cm}
\begin{center}
Deriving Veneziano Model in a Novel 
String Field Theory Solving String Theory 
by Liberating Right and Left Movers
\end{center}
\end{Large}

\begin{center}
{\large Holger Bech Nielsen$^{a)}$ 
and Masao Ninomiya$^{b)}$
}

\vspace{0.4cm}

a)
The Niels 
Bohr Institute,\\ Blegdamsvej 17 - 21, 
Copenhagen {\O}; Denmark ,\\ 
E-mail: 
hbech@nbi.dk, 
hbechnbi@gmail.com\\

b)
Okayama Institute for Quantum Physics,\\
Kyoyama 1-9-1 Kita-ku, Okayama-city 
700-0015, Japan\\
E-mail: msninomiya@gmail.com

\end{center}

\begin{center}
PACS NO: 11.25.-w, 11.27.+d, 11.10.-z, 03.70.+k, 11.25.Wx

\end{center}

\begin{abstract}

Bosonic string theory with the possibility
for an arbitrary number of strings - i.e.
a string field theory - is formulated by
a Hilbert space (a Fock space), which is 
just that for massless noninteracting 
scalars. We earlier presented this novel 
type of string field theory, but now we 
show that it leads to scattering just 
given by the Veneziano model amplitude.
Generalization to strings with fermion 
modes would presumably be rather easy.
It is characteristic for our formulation
/model that:
1) We have thrown away some null set of 
information compared to usual string field theory, 2)Formulated in terms of our 
``objects'' (= the non-interacting 
scalars) there is no interaction and 
essentially no time development(Heisenberg
picture), 3) so that the S-matrix is 
in our Hilbert space given as the unit 
matrix, S=1, and 4) the Veneziano 
scattering amplitude appear as the overlap
between the initial and the final state 
described in terms of the ``objects''.
5) The integration in the Euler beta 
function making up the Veneziano model 
appear from the summation over the number 
of ``objects'' from one of the incoming 
strings which goes into a certain one of 
the two outgoing strings.
Due to a correction needed to fit the Veneziano 
model form from the Weyl anomaly in 2 dimensions 
we have to have the dimension 26 as usually required for 
the bosonic string.   
\end{abstract}

\begin{center}
{\em This work is to be published in the Proceedings of 
the 17th Bled Workshop on ``What Comes 
Beyond the Standard Models'' in 2014 
organized by Norma Mankoc-Borstnik, Dragan 
Lukman, M. Khlopov, and H.B. Nielsen}  
\end{center}

\section{Introduction\ldots}
\label{s:intro}
We have already earlier put forward 
\cite{1,2,Novel} ideas towards a novel 
string field theory (meaning a second 
quantized theory of strings from string
theory \cite{12, 13, 14, 15, 17}), which
of course means the theory\cite{3,4,5,6,7,8,9,10,11} in which  
you can describe several strings like one 
describes several particles at a time in 
in quantum field theory.
It may be understood that we as other 
string field theories in or theory have
a Hilbert space or Fock space the vectors 
of which describe states of the whole 
universe in the string theory. Our model
is similar to the formulation of Thorn
\cite{new12} and also use a discretization
like ourselves, but we discretize after 
having separated right and left movers as 
we shall see below.   This Hilbert space, 
which describes the states of the universe,
turns in our model/formulation out to be 
really surprisingly simple in as far 
as it is simply the second quantized Fock
space of a non-interacting massless scalar
25 +1 dimensional particle theory in the 
bosonic string case!

Let it be immediately be stated that 
although our formulation/model is supposed
just a rewriting of string theory - and 
thus in its goal there is nothing new 
fundamentally - it is definitely new 
because we throw away compared to usual
string theory and usual string field 
theory as Kakku and Kikkawa's and Witten's
a {\em null set of information}. The 
information, which we throw away is the 
one about how the different pieces
of strings hang together. That is to say
we rather only keep the information about 
where in target space time you will find a
string and where not. Due to this throwing
away of information and other technically 
doubtful treatment of the string theory 
by us it is a priori no longer guaranteed
that our string field theory appearing 
as just the non-interacting massless scalar
theory in 25+1 dimensions is indeed just 
a rewriting of string theory. Rather one 
should see our progresses such 
as the derivation of the string spectrum 
\cite{Novel} in reproducing 
usual properties of string theory from
our model/formulation as tests  
that indeed our model is in spite of 
the null set of information thrown away
indeed the full string theory.

 The major 
achievement in the present work is 
also such a test, namely testing that 
our model/formulation leads to the 
Veneziano model scattering amplitude for
scattering of strings formulated in our 
novel string field theory. 
  
The particles that formally occurs in the 
construction of our Hilbert space or 
Fock space of our model or formulation 
of string theory we call ``even objects''
and each such ``even object'' has in our 
formulation a kind of momentum variable 
set $J^{\mu}$ ( it is proportional to 
a contribution to the total momentum 
of the string to which it belongs).
Really this $J^{\mu}$ has as some technical
details got its longitudinal momentum
(in target space time of 25 +1 dimensions) 
component $J^+ = J^0 + J^{25}$ fixed by 
what corresponds to a gauge choice in the
string parameterization to be 
\begin{equation}
J^+ = \frac{a\alpha'}{2},
\label{gaugefixing}
\end{equation}  
(We shall below that we end up being 
driven to also allow 
$J^+ = -\frac{a\alpha'}{2}$)
and its infinite momentum frame energy
proportional component $J^- = J^0 - J^{25}$
is written just by the mathematical 
expression ensuring the light-likeness  
\begin{equation}
(J^{\mu})^2 = \eta_{\mu\nu}J^{\mu}J^{\nu}=0
\label{lightlike}
\end{equation}  
of the ``even object'' momentum-like
$J^{\mu}$ variable. Thus the only genuine
degrees of freedom components of this 
even object variables $J^{\mu}$ are the 
``transverse'' components corresponding 
to the first 24 components, namely those 
having $\mu = i$ where $i = 1,2,3,...,24$,
i.e. $J^i$. In addition the 
``even objects'' have 24 conjugate 
momenta $\Pi^i$, conjugate to the $J^i$'s,
so that 
\begin{equation}
[\Pi^i, J^j] =i \delta_{ij}
\end{equation}     
for $\Pi^i$ and $J^j$ belonging to the 
same even object of course.

Our Hilbert space for states of the 
universe corresponds now simply to a set
of harmonic oscillators, one for every set
of  $J^i$-value combinations (of 24 real
numbers), and the creation operator for 
an ``even object'' with its $J^i$'s being
$J^i$ is denoted $a^{\dagger}(J^i)$. Since 
there is a calculational relation between
the set $J^i$ of the transverse 
components and the full 26-vector 
$J^{\mu}$ given by adding the equations
(\ref{gaugefixing},
\ref{lightlike}), we could  
equally well use as  the symbol in the 
creation and annihilation operators 
$J^{\mu}$ as the symbol $J^i$, and so we 
have by just allowing both notations  
$a^{\dagger}(J^i) = a^{\dagger}(J^{\mu})$, where it is understood that the $J^{\mu}$ is 
calculated from the only important 
transverse components $J^i$. Similarly 
the destruction operators are $a(J^{\mu})
= a(J^i)$ and we shall think of the 
Hilbert space describing the states of the 
Universe (in a string theory world) as 
having basis vectors of the type 
\begin{equation}
a^{\dagger}(J^i(1)) a^{\dagger}(J^i(2))\cdots
a^{\dagger}(J^i(L))|0>.
\end{equation}   
To tell the truth we though better reveal
the little technical detail, that this 
simple situation with only one type of 
`` even objects'' that can exist in the 
states described by $(J^i, \Pi^i)$ is only
true for the case of a string theory model 
{\em with open strings}, while we for the 
case of a string theory with only closed 
strings must have {\em two kind of even objects} that can be put into the 24 or 
26=25+1 dimensional (Minkowski) space, one
right denoted by R and one left denoted 
by L. So in the only closed string case 
we could even naturally consider it that 
the two types of even objects ``live'' in
two different Minkowski spaces - one R and one L-. The figure \ref{fig1} illustrates these two 
slightly different cases.     
\begin{center}  
\includegraphics[clip]{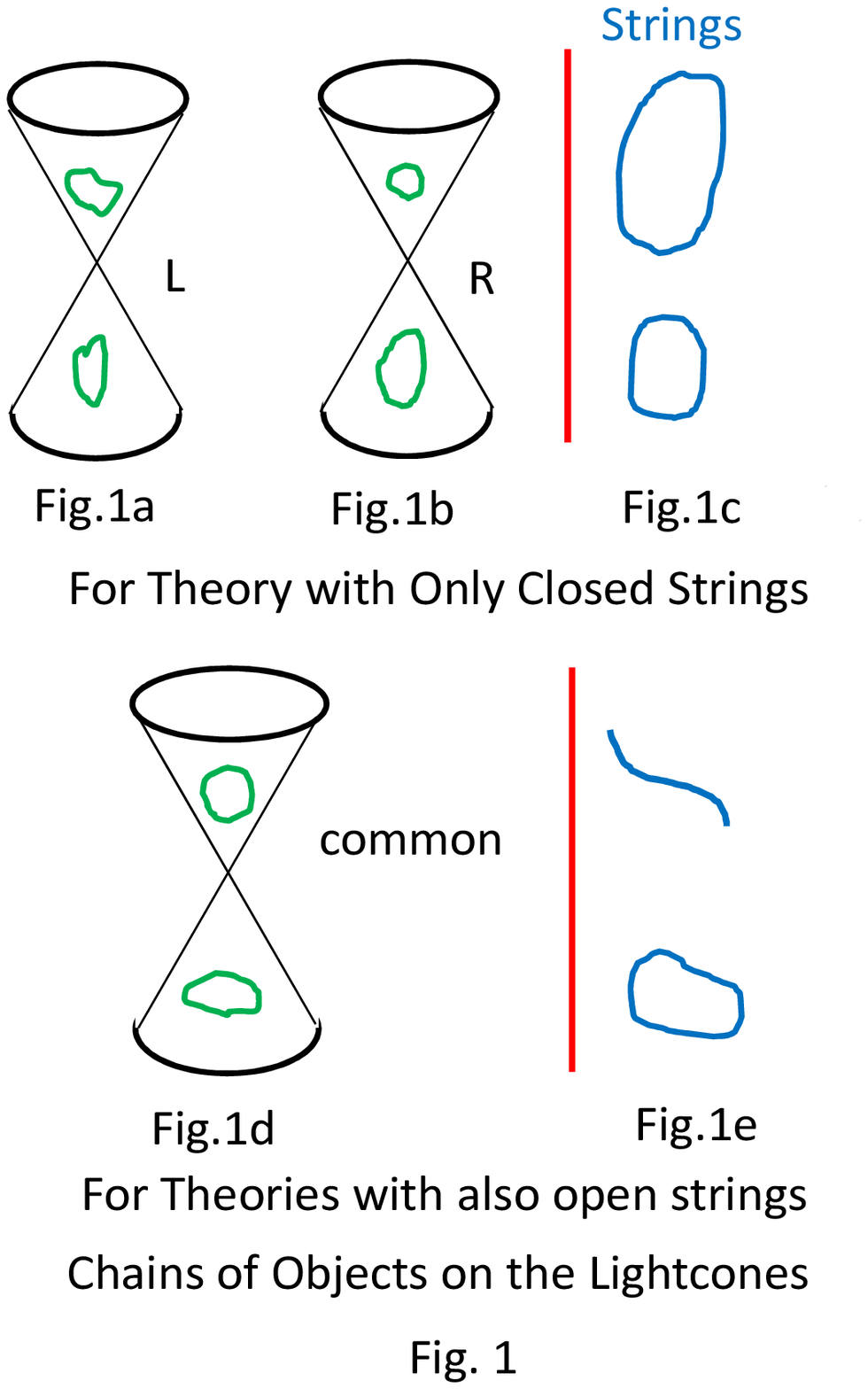} 
\label{fig1}
\end{center}
The connection between the strings present 
in a given state of the universe and the 
even objects corresponding to a set of 
strings is a priori not completely trivial
and has to be described. It is not 
hundred percent true that the strings 
consist of even objects, but there is so 
much about it that there is an actually
infinite number of even objects 
corresponding to each string present.
This divergent number of many even 
objects in a string is given as a function
of the small parameter $a$  already 
mentioned in formula (\ref{gaugefixing}).

\section{Correspondence from 
Strings to Objects}
The crux of the matter in the formulation of our string 
field theory model or formulation is  
to put forward the rule for how a 
given string state is translated into a 
state described in terms of a state of 
what we call ``even objects'': 

In the case of a theory with only closed 
strings we shall make use of the {\em 
solution in the conformal gauge for the 
26-position fields $X^{\mu}(\sigma, \tau)$
in terms of right and left movers}. 
Remember that the time-development of a 
string in string theory is described by 
letting its timetrack - which of course 
becomes a two-dimensional surface in the 
25+1 dimensional ``target space'' - be 
parameterized by the two real variables 
called $\sigma$ and $\tau$. At first 
one may think of these parameters as 
parameterizing the timetrack surface in an
arbitrary way and therefore one even has 
to have an action - the Nambu(-Goto) 
action - chosen so as to be invariant 
under reparameterization, meaning that 
one goes over to a new set of coordinates
parameterizing the timetrack $(\sigma', 
\tau') = (\sigma'(\sigma,\tau), 
\tau'(\sigma,\tau))$. This requirement 
of reparametrization invariance fixes up
to an overall constant the action to be
given by the area of the timetrack 
surface
\begin{equation}
\hbox{Single string action}= S_{Nambu}
\propto \hbox{area} = \int \sqrt {
det \left ( \begin{tabular}{c|c}
$(\dot{X}^{\mu})^2$ & 
$\dot{X}^{\mu}\cdot X'^{\mu}$\\
$ \dot{X}^{\mu}\cdot X'^{\mu}$& $(X'^{\mu})^2$
\end{tabular} \right ) }d\sigma d\tau,
\label{Nambu}
\end{equation} 
where we have as usual denoted
\begin{eqnarray}
\dot{X}^{\mu}(\sigma,\tau)& 
\stackrel{def}{=}& \frac{
\partial X^{\mu}(\sigma,\tau)}{\partial 
\tau}\\
X'^{\mu}(\sigma,\tau)& \stackrel{def}{=}&
\frac{\partial X^{\mu}(\sigma,\tau)}
{\partial \sigma}.
\end{eqnarray}
For an open string the timetrack is like 
a band extending in the time direction,
while for a closed string the track is topologically like a tube/cylinder also  
extending roughly in time 
direction.

Now one usually in steps fix the ``gauge''
meaning the parameterization, i.e. the 
choice of a new set of coordinates which
we again may call $(\sigma, \tau)$(leaving 
out the prime on $(\sigma',\tau')$). 
The first step in the gauge choosing is 
what is called conformal gauge choice and 
corresponds to arranging the coordinate 
equal constant curves to be orthogonal 
seen from the external/target space of 
25+1 dimensions. Often in literature 
one works with Euclideanized $\sigma$ and
$\tau$ as if the string timetrack were 
a two dimensional Euclidean space, but 
thinking physically on a true string the 
space felt by a being living 
attached onto  the string would  
be a 1+1 dimensional space time 
with one time dimension and one spatial 
dimension. For the thinking of the 
present article and our foregoing works 
on our novel string field theory we shall
take this latter - more physical - point
of view that the internal space time is
indeed a space-{\em time}. We think of
$\tau$ as the time coordinate and of 
$\sigma$ as the spatial coordinate 
along the string.

After having chosen the ``conformal 
gauge'' the equation of motion derived 
from the Nambu action at first simplifies 
and together with the constraints 
appearing due to the reparameterization 
symmetry of the original action we can 
summarize the equations in the conformal
gauge:
 \begin{eqnarray}
\square X^{\mu}(\sigma,\tau) &=&0 (\hbox{equation of motion)}\\
(\dot{X}^{\mu}(\sigma,\tau))^2 
-(X'^{\mu}(\sigma,\tau))^2 &=&0\hbox{
(constraint)}\\
\dot{ X}^{\mu}(\sigma,\tau)\cdot X'^{\mu}
(\sigma,\tau) &=&0\hbox{(constraint also)}.
 \end{eqnarray}

\begin{center}
\includegraphics[clip]{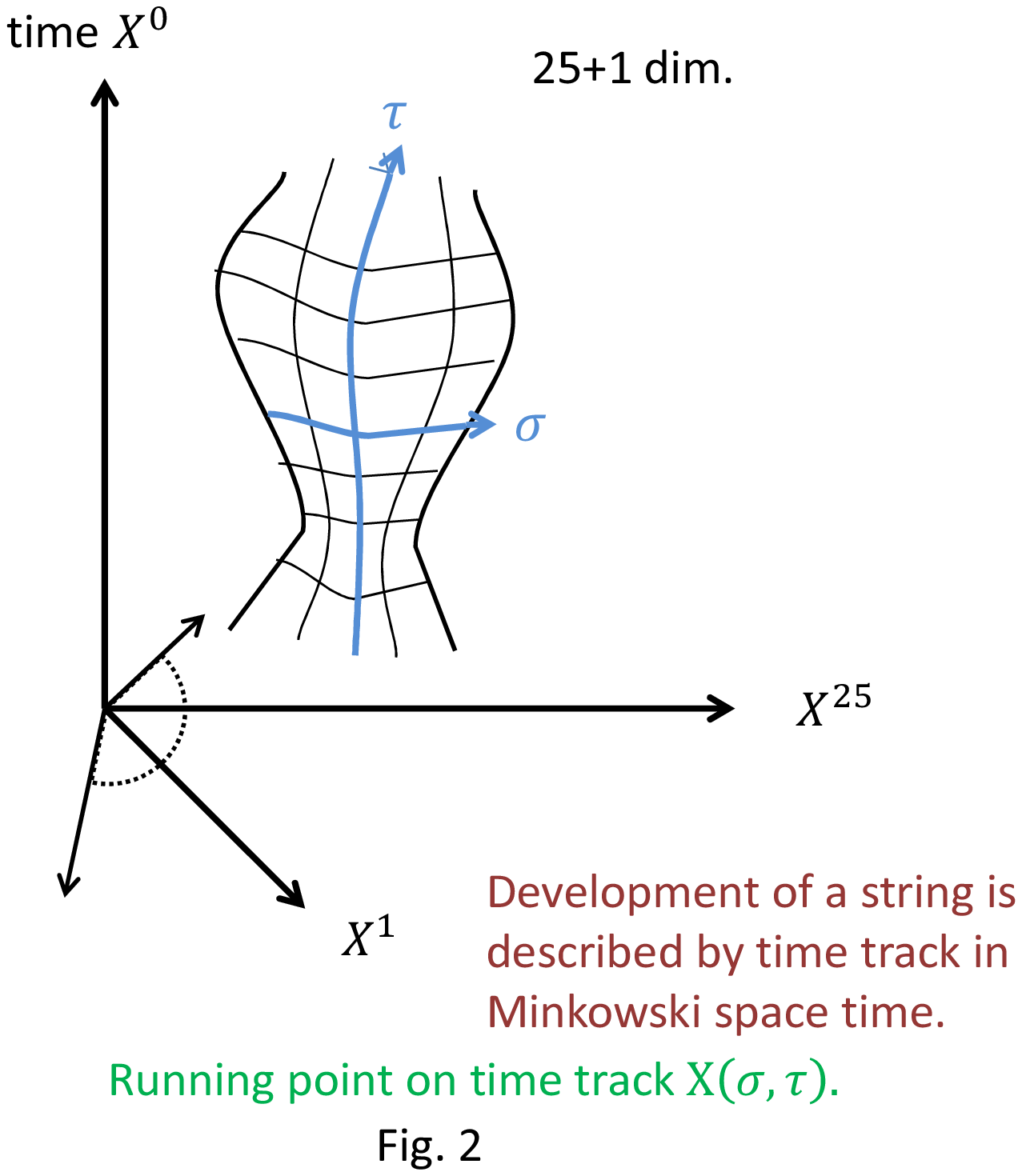}
\end{center}
Here the D'Alambertian 
\begin{eqnarray}
\square &=& \partial_{\tau}^2 - \partial_{\sigma}^2 = (\partial_{\tau}-\partial_{\sigma})
(\partial_{\tau}+\partial_{\sigma}),
 \end{eqnarray}
and the eqations of motion are easily 
{\em solved} by the ansatz
\begin{equation}
X^{\mu}(\sigma,\tau) = X_R^{\mu}(\tau-\sigma)
+ X_L^{\mu}(\tau+\sigma), 
\end{equation}
which is importance for our 
novel string field theory in as far as 
it is actually the $\tau$-derivatives
of the 26-vectorial functions in the 
solution $X_R^{\mu}(\tau-\sigma)$ and 
$X_L^{\mu}(\tau+\sigma)$, which are going to
be identified as we shall see soon by our 
``objects''. Note immediately, that 
these right and left mover variables 
$X_R^{\mu}$ and $X_L^{\mu}$ only depend 
on {\em one} variable each, namely 
respectively on $\tau_R \stackrel{def}{=} 
\tau -\sigma$ and $\tau_L\stackrel{def}{=}
\tau +\sigma$, so that the equations of 
motion with $\tau$ conceived of as the 
time have indeed been solved. The ansatz 
functions $X_R^{\mu}$ and $X_L^{\mu}$ are 
more like initial conditions for the 
solution. 

In terms of these initial condition 
variables  $X_R^{\mu}$ and $X_L^{\mu}$
the constraints take the very simple form
\begin{eqnarray}
(\dot{X}_R^{\mu}(\tau_R))^2=
(\dot{X}_R^{\mu}(\tau-\sigma))^2&=&0
\hbox{(constraint)}\\
(\dot{X}_R^{\mu}(\tau_L))^2=
(\dot{X}_R^{\mu}(\tau+\sigma))^2&=&0
\hbox{(constraint)}
\label{constraints}
\end{eqnarray}

The overview of description of our object
rewriting of the string theory is that 
we let there be an object for every point 
in (a period for) the coordinates $\tau_R$ and $\tau_L$
in the case of only \underline{closed} strings, and 
that the objects are closely related to 
the variables $\dot{X}_R^{\mu}$ and 
$\dot{X}_L^{\mu}$. For continuity of these
variables as  functions of respectively
$\tau_R$ and $\tau_L$ the images of these
functions $\dot{X}_R^{\mu}$ and   
 $\dot{X}_L^{\mu}$ are - except 
for fluctuations at least - smooth curves,
because of the constraints (\ref{constraints}) these curves must 
lie on the lightcone(s). 

\begin{center}
\includegraphics[clip,scale=1.5]{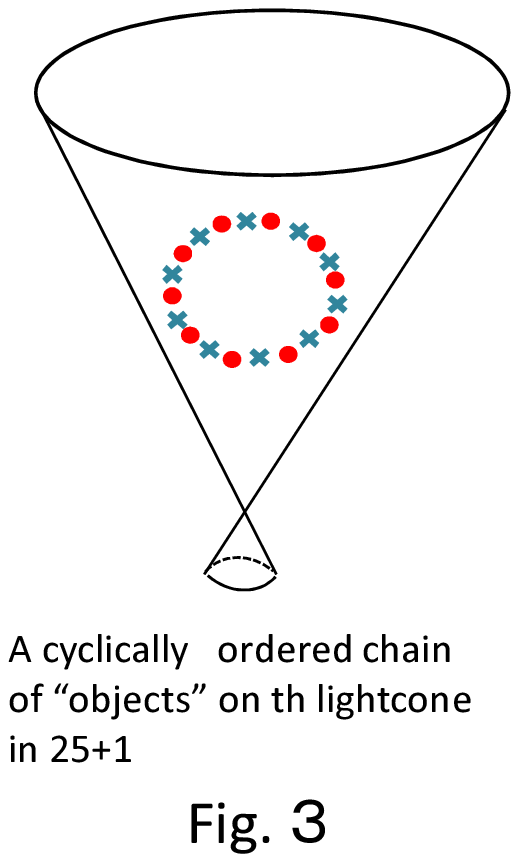}
\end{center}
Since we can also
consider the variables $\tau_R$ and 
$\tau_L$ as $\sigma$-variables for 
constant $\tau$ the periodicity w.r.t.
$\sigma$ of the position variables etc. - 
for in fact both open and closed strings- 
but at least clearly for the closed 
strings, comes to imply that the just 
mentioned images for $\dot{X}_R^{\mu}$ 
and  $\dot{X}_L^{\mu}$ become closed curves
on the light cone.


Two points further are illustrated by the 
figure 3: 1) We ``discretize'', so that we 
replace the in principle continuum 
infinity of $\tau_R$ or $\tau_L$ values 
by a series of discrete points with a 
``distance between these points'' being 
proportional to a small quantity $a$ 
later taken to go to zero. 2) We treat the 
even numbered and the odd numbered 
``discretized points'' differently, as
is on the figure illustrated by them 
being denoted differently by dots and 
crosses.

\subsection{Open String Case}
The open string case has a tiny technical
complication: 

At the boundaries of the open string one 
has boundary conditions which are translated 
into our model favorite language 
of $\dot{X}_R^{\mu}$ and $\dot{X}_L^{\mu}$
implies
\begin{eqnarray}
\dot{X}_L^{\mu}(\tau) &=&\dot{X}_R^{\mu}
(\tau) \hbox{from boundary at $\sigma 
=0$}\nonumber \\
\dot{X}_L^{\mu}(\tau) &=&\dot{X}_R^{\mu}(
\tau - 2\pi) \hbox{from the $\sigma=\pi$ 
end boundary.}, \label{boundaries} 
\end{eqnarray}
where we assumed the notation that the 
length of the string in $\sigma$-parameter
language is $\pi$. These two boundary 
conditions (\ref{boundaries}) imply that 
we can identify, for the open string, 
the $\dot{X}_R = \dot{X}_L$ and have that
this common (differentiated)``initial 
condition variable'' must be periodic with
the period $2\pi$ meaning twice 
the $\sigma$-variable range corresponding 
to the string length. Thus in the beginning announced we got 
the fact 
that while we for only closed string 
theories have to distinguish $\dot{X}_R$ 
and $\dot{X}_L$, this is no longer needed
for an open string.

\subsection{More Precise 
Correspondence between Strings 
and ``even objects''}

More precisely we shall divide up 
into ``discretized'' pieces the $\sigma$ 
range around a closed string or the tour 
forward and backward along an open string
into, let us say, $N$ pieces. What we really
want, is to divide up a period for 
say $\dot{X}_R^{\mu}$ in its argument
$\tau_R$ (and in the closed case the same 
for $\dot{X}_L$, while in the open string 
case just identify the left and right 
mover variables because of boundary 
conditions). The precise way of dividing 
up could be thought of as dividing in 
equal steps in the variable, say $\tau_R$,
but there is still some coordinate 
specification/gauge choice left even after
the conformal gauge choice. In fact 
one can still as such a rudimentary 
freedom of choosing coordinates select
any (increasing) function $\tau_R'(\tau_R)$
and any (increasing) function 
$\tau_L'(\tau_L)$ as a new set of 
coordinates (having in the background of the mind  the
identifications $\tau'= \tau -\sigma'$ and
$\tau_L = \tau+\sigma$). By discretizing 
we replace essentially a variable as 
$\tau_R$ by an integer valued variable 
- counted modulo $N$ if $N$ corresponds 
to the period -, so that the $\tau_R $
value corresponding to the integer $I$
is denoted $\tau_R(I)$. Interpolating
we can easily make an approximate sense 
of even $\tau_R$ defined for non-integer 
values of $I$. Thus we formally associate
any string with a series something we call
``objects''- and which is something only 
defined in our model -, which are 
characterized each by a set of degrees 
of freedom (as if it were particles):
$J_R^{\mu}$ and the conjugate variables
$\Pi^{\mu}$ or better only $\Pi^i$ where 
the $i$ only runs over the transverse 
coordinates $i=1,2,...,24$. The reader 
may crudely think of these objects as a 
kind of partons, but really we simply 
have to define them by their relation to 
the $\dot{X}_R^{\mu}$ (and for the closed 
string case also to the separate 
$\dot{X}_L^{\mu})$. To every discretization 
point on the $\tau_R$ axis say, let us say
discretization point number $I$, we 
associate an ``object'' for which the 
dynamical variables $J_R^{\mu}(I)$ are 
given 
as 
\begin{eqnarray}
J_R^{\mu}(I)& =& X_R^{\mu}(\tau_R(I+1/2))
-X_R^{\mu}(\tau_R(I-1/2)). \label{Jdef}
\end{eqnarray}     
Notice that since the difference between 
the two argument values $\tau_R(I-1/2) $
and $\tau_R(I+1/2)$ is small this 
definition of $J_R^{\mu}(I) $ for a 
discretization point on the $\tau_R$-axis
in reality means that 
\begin{equation}
J_R^{\mu}(I) \approx \dot{X}_R^{\mu}\frac
{d\tau_R}{dI},
\end{equation}
and so indeed as announced our variables 
$J_R^{\mu}$ assigned to the ``objects''
are ``essentially'' the 
$\tau_R$-derivative $\dot{X}_R^{\mu}$ of
the right mover $X_R$ part of the solution.

\subsection{The Even Odd Detail}
Now there is an important technical 
detail in the setup of our model:

We have the problem that if one shall
make creation and annihilation operators
for some ``objects'' in a way analogous 
to how one in usual quantum field theory
have creation and annihilation operators 
for particles, one shall describe these 
creation and annihilation operators by 
having an argument for a set of variable 
describing the ``object'', a set of 
variables which {\em commute with each 
other}. It is indeed well-known that one 
must in quantum field theory {\em either }
take the creation and annihilation 
operators to be functions of the spatial
momenta of the particles 
created/annihilated {\em or} one can use 
instead position variables, and that 
corresponds to working with the second 
quantized fields ${\cal \phi}(x)$. But 
the usual simple mutual commutation rules 
for creation and annihilation operators 
could not be obtained if one would 
attempt to construct them to correspond to a combination of dynamical variables for 
the particles that did not commute with 
each other. What could also a creation 
operator depending on mutually 
complementary variables for a single 
particle correspond to creating ?
It could not create a particle with the 
specified quantum numbers in such a case 
because that would be against the 
Heisenberg uncertainty relation. In the 
corresponding way we must choose whatever 
variables we let our ``objects'', to be 
associated with creation, and annihilation 
operators depend on be arranged so as to 
commute. But then we have problem, because
does our $\dot{X}_R^{\mu}$ 's which are 
proportional to the object-variables 
$J_R^{\mu}$ commuting? No, they do not 
commute in as far as the theory of a 
single starting from the Nambu Lagrancian 
e.g. in the conformal gauge leads to
\begin{equation}
[\dot{X}_R^{\mu}(\tau_R'),\dot{X}_R^{\nu}
(\tau_R)] \propto g^{\mu\nu}\delta'(\tau_R'
-\tau_R).\label{commutationXdot}
\end{equation}   
Thinking discretizing, such a derivative 
of a delta-function commutator means that 
in the discretized chain the $\dot{X_R} $
or equivalently $J_R$ 's which are {\em 
next neighbors do NOT commute}.

So we had to invent a trick to avoid to
have to make creation and annihilation
operators for ``objects'' sitting in the 
chains of ``objects'' along the variable 
$\tau_R$ as neighbors in the 
discretization. 

The trick which we have chosen consists 
in {\em only using in the creation and 
annihilation operators every second of
the by discretization by (\ref{Jdef})
defined ``object''-variables $J_R$.}
That is to say we choose to {\em only
construct creation and annihilation 
operators} for those ``objects'', which 
in the discretized series of objects 
for a given string have got an {\em even 
number $I$}. One could say that we in our
model construct our Hilbert space only 
in terms of such ``even objects'', and 
one could almost say only consider these 
``even objects'' as ``really existing''
in our basic Hilbert space description.

But then we are coming to the 
problem that we need to a full 
description of the string states also the 
``odd objects'': what to do about them?
We say that when you have a series of the 
``even objects'' on a string, we make 
the rule to construct in between any two 
next to neighboring ``even objects''
(i.e. two ``even objects '' deviating
in number by just 2) an ``odd object'' 
from the conjugate momenta $\Pi^i$ say 
of the neighboring ``even objects''.
( There is another technical detail 
connected with the + and - components in
the infinite momentum frame we have chosen
to work with, so we shall avoid 
discussing conjugate momenta to other than
the transverse components $J_R^i$ - the 
first 24 components -. Therefore we 
only consider these first 24 components 
of conjugate momenta to the $J_R$'s).
In fact we have to take the following 
rule for constructing the ``odd object''
$J_R^i$ components for the ``odd object''
number $I$ (where $I$ then is an odd 
integer (modulo the even number $N$)),
\begin{eqnarray}
J_R^i = -\pi \alpha' (\Pi_R^i(I+1) -
\Pi_R(I-1)),\label{oddfromeven} 
\end{eqnarray}        
 in order to obtain the commutation 
rule corresponding to the derivative 
of delta function commutation rule
(\ref{commutationXdot}) discretized.

The reader should check and understand 
that with this construction of the 
``odd objects'' any quantum state of the 
string expressed as a state of the 
variables $\dot{X}_R^i$ ( and for 
the closed string also $\dot{X}_R^i$)
can be expressed as a corresponding 
quantum state of a set of $N/2$ ( $N$ must be even) ``even objects'', because the 
even object commutation rules 
\begin{equation}
[J_R^i(I),\Pi_R^k(K)] = i\delta^{ik}
\delta_{IK} 
\end{equation}
corresponds just to the commutation rules 
for the $\dot{X}_R^i$ ( and $\dot{X}_L^i$).
There is though one little technical 
detail to be studied in later works:
The absolute position of the string were 
differentiated away from our discussion 
by dotting the $X_R$ and $X_L$and 
correspondingly the formula for the 
``odd objects'' does {\em not} make use 
of the sum over all the ``even object''
conjugate variables $\Pi_R^i$ around the 
closed chain. So there is suggestively the 
possibility of identifying the average 
position of the string 
proportional to this sum over all the 
conjugate to even object variables.
   
\subsection{Several Strings}
So far we should have now given the 
prescription 
for constructing a cyclically ordered 
chain of ``even objects'' corresponding 
to a given quantum state of a single open 
string. (If one wants a closed string 
one shall construct two cyclically ordered 
chains of ``even objects'' one for
right movers consisting of $J_R$ ``even 
objects'' and consisting $J_L$ left mover
``even objects''). Since the commutators 
were arranged to isomorphic to the 
discretized $\dot{X}_R$ 
( and $\dot{X}_L$)
it should be possible to construct such 
an ``even object'' state. It is then 
of course also trivial and completely 
analogous to usual quantum field theory
construction of a state with $N/2$ 
particles to construct a Hilbert space 
( Fock space) state for $N/2$ of our 
``even objects''. Corresponding to 
a single open string we thus simply 
have Hilbert space state with the large
(divergent in the limit of 
$a\rightarrow 0$)
infinitely many ($=N/2$) ``even objects''
sitting approximately in a cyclic chain 
on the light cone in a 25 +1 dimensional 
(Minkowskian) $J_R^{\mu}$-space. 

But now it is the main point of a model
being a string field theory (SFT), that 
such a model {\em can describe several 
strings in one Hilbert space state}.
Once we have made our formulation of one 
string in our ``even object'' formulation
it is, however, rather trivial to 
construct states with an arbitrary number 
of strings. One can just act on the 
``zero even object'' with all the product 
of creation operators corresponding to 
the various strings - we want to have in 
the state to be described - each creating 
the ``even objects'' associated with 
the string in question.    
So to speak if string number 1 is 
described by the product 
\begin{eqnarray}
C_1|0>&=& \int \Psi_1 (J_R^I(0), 
J_R^i(2), ...
, J_R^i(N-2))\cdot\nonumber \\
&&\Pi_{I = 0,2,4,...,N-2}
 a^{\dagger}(J_R^i(I))\nonumber\\
&&\cdot\Pi_{I=0,2,...,N-2}
\Pi_{i=1,2,...,24}dJ_R^i(I)|0>,
\label{stateC_1}
\end{eqnarray}
where $\Psi_1 (J_R^I(0), J_R^i(2), ...
, J_R^i(N-2))$ is the wave function for the
state of the single string 1 described 
in terms of  ``even objects'',
and the string number 2 by an analogous 
expression, then a state with both string 
1 and string 2 (say they are open) is 
given as
\begin{eqnarray}
C_1 C_2 |0>& =& \int \Psi_1 (J_R^I(0), J_R^i(2), ...
, J_R^i(N-2))\nonumber \\
&&\Pi_{I = 0,2,4,...,N-2}
 a^{\dagger}(J_R^i(I))\Pi_{I=0,2,...,N-2}
\Pi_{i=1,2,...,24}dJ_R^i(I)\nonumber\\
&&
\cdot\int \Psi_2 (J_R^I(0), J_R^i(2), ...
, J_R^i(N-2))\Pi_{I = 0,2,4,...,N-2}
 a^{\dagger}(J_R^i(I))\nonumber\\
&&\Pi_{I=0,2,...,N-2}
\Pi_{i=1,2,...,24}dJ_R^i(I)|0>.
\end{eqnarray}
Luckily the commutation of the creation 
operators for ``even objects'' makes it 
unnecessary to specify any order in which
the creation operator products 
corresponding to different strings have 
to be written.

We thus have a scheme for constructing 
Hilbert space states - in the Hilbert 
space which is really that of massless 
scalar ``even objects'' conceived of as
particles in an ordinary Fock space - 
corresponding to any number of strings 
wanted. In this sense we have a string 
field theory.

\subsection{Final Bit of Gauge Choice}
As already mentioned the choice of 
parameterization (often called gauge choice)
were not finished by the conformal gauge, 
since we could still transform the 
variables $\tau_R$ and $\tau_L$ by 
replacing them by some 
increasing function of themselves. 
In the ``infinite momentum frame gauge''
the choice is to fix this freedom 
by requiring the density of $P^+
=P^0+P^{25}$ (longitudinal momentum
we can say) momentum is constantly measured 
in say $\tau_R = \tau - \sigma$ or say
in $\sigma$. When we discretize as 
described above and take it that the 
$\tau_R$ distance between neighboring 
``objects'' along the $\tau_R$-axis 
should be the same all along, then this
gauge choice comes to mean that each 
object gets the same $P^+$ momentum.
We can therefore describe this gauge 
choice - which is essentially the usual 
one in infinite momentum frame, just 
discretized our way  - by saying that we 
impose 
some fixed small value for the $+$ 
component of our ``object''(-variables)
\begin{equation}
J_R^+ = \frac{a \alpha'}{2} \hbox{ in our 
first attempt (later problem comes)}.
\end{equation} 
With this gauge choice we have made the 
number of objects $N$ and thus of ``even
objects'' $N/2$ proportional to the $P^+
=P^0+P^{25}$
component of the 26-momentum of the string
in question. So e.g. the conservation of
this component of momentum corresponds to 
the conservation of the number of say 
``even objects''. After this choice of 
gauge extremely little is left to be 
chosen for the reparameterization: you can still for the closed string shift the 
starting point called $\sigma=0$, but 
that is all. Corresponding to this 
extremely little reparameterization left 
unfixed you can still cyclically shift 
along the topological circles on which 
the objects of a string sits, and that 
turn out due to the possibility for adding a constant to $\tau$ also to be true for 
the open string. The objects corresponding 
to a cycle for a string are cyclically 
order but the starting point is still 
an unchosen gauge ambiguity. To an open 
string we have one such loop or cycle, and
to a closed one we have two.

\section{Comparing Our String 
Field Theory
to Other Ones}

It should be stressed that our ``novel''
string field theory really is novel/new, 
since it deviates from earlier ones like 
Kakku and Kikkawa or Wittens string 
field theories in important ways even 
if some calculations should soon turn out similar:
\begin{itemize}
\item{1.} The information kept in our 
formalism is {\em not} the full one kept 
by the theories by Kaku Kikkawa or by 
Witten, but deviates by having relative 
to these other string field theories 
thrown out  - actually only a null set 
of - information. It is the 
information about how the strings hang 
together, that is thrown out. We could say
that we - Ninomiya and Nielsen - only in
our rewritten string states keep track of
where in the space time you may see a 
piece of string, but not of how one piece
hangs together with another piece. If a 
couple of strings cross each other there 
is a point in target space wherein four 
pieces of string meet, two belonging to 
each of the crossing strings. In usual 
string field theories, such as Kaku 
and Kikkawa \cite{3} and Witten's\cite{4}, it is part of 
the information kept in the Hilbert space 
vector describing the state of the 
universe which of these 4 pieces are 
connected to which. In our formulation,
however, {\em this information has been 
dropped}.
         
\item{2.} A further consequence of this 
drop of information is that if two strings 
scattering by just exchanging tails - as 
one must think scattering should typically
happen classically - then really nothing
have to happen at all in our formalism.

Indeed it is a second characteristic 
property of
our string field theory model, that 
in the scattering counted in terms of our 
``even objects'' (which are the ones
truly represented in the Hilbert space;
the odd ones are just mathematical 
constructions from the conjugate variables
for the ``even'' ones) {\em nothing 
happens!} The scattering process is not
represented in the Hilbert space formalism
of ours.   
  \item{3.} A consequence of item{2.} is 
that the S-matrix gets calculated formally
as an {\em overlap} of the initial with
the final state.
\item{4.} And this fact is also connected 
with that the Hilbert space or Fock space
of our formulation is the extremely simple
free massless scalar Fock space. Actually
though there is gauge fixing, that makes 
the states of the ``even objects'' even 
have their $J^+$ components fixed by 
(\ref{gaugefixing}). This is contrary to 
the other string field theories which have 
much more complicated structures.
\item{5.} But perhaps the most important 
distinction for the other string field 
theories is that {\em we use a description
in terms of something quite different from
the strings themselves, namely our 
`even objects''}, while the other string 
field theories have quite clearly all 
through their formalism the strings one 
started from. In ours the string has been
hit to the extent that we at the end 
must ask: What happened to the string?
The answer is roughly that there is no 
string sign left in the Hilbert space 
structure of being only that of free 
massless scalars. Rather {\em the string 
in our formalism only finds way into the 
calculations via the initial and final
states put in!} That is to say that in 
our formalism it looks that the whole 
story of the strings only will appear 
because there is an extra ``stringy''
assumption put in about the initial
state - and presumably it is necessary 
even to put it in for the final state -
so that the whole string story is not part of the structure of the theory nor of the 
equation of motion, but rather on an equal
level with the cosmological start of the 
Universe, or the initial conditions of 
low entropy allowing there to be a second 
law of thermodynamics. If it should turn
out that indeed even extra assumptions 
about {\em the final states} are needed to
make our formalism function as a string
theory, then one could say that in our 
formalism an influence from future is 
required.   
\end{itemize}  
 
With all these deviations from the usual 
string field theories, one may
worry whether our rewriting truly is a 
rewriting and thus can count as a true 
string field theory, because does it 
indeed describe the conventional string 
theory, or could it be that we had thrown
away too much (even though only a null 
set)?

Because of this possibility that our model
does not truly represent string theory at 
the end it becomes important - also for
the purpose of testing if our model is 
string theory - to check the various 
wellknown features of string theory. 
We have not long ago published an article 
\cite{Novel} in which we showed that the 
mass spectrum of the strings in our 
string field theory became the usual one.
This is one such little check that our 
model/string field theory is on the right 
track. In the succession of this article 
we shall concentrate on sketching the 
calculation of the scattering amplitude 
for two ground state strings (tachyons)
scattering elastically into two also
tachyonic ground state open strings.

Actually it turned out that we were not 
quite right in the first run, because we 
only get one term out of three terms 
that should be present in the correct 
Veneziano model. This little shock we 
sought to repair by modifying our gauge 
fixing condition and allowing ``even objects'' also with negative $J^+$. As we 
shall see later we think it reflects a 
more general problem with infinite 
momentum frame. 

\section{Yet More Technical 
Details}     
\subsection{The + and $-$ 
components of $J_R$}
Especially if one wants to get an idea 
about our work \cite{Novel} checking the 
spectrum of our strings it is necessary to
keep in mind that it is only the 
components $J_R^i$ for $i=1,2,...,24$, 
which are simply independent dynamical 
variables for the ``even object''. The 
remaining two components are not 
independent. Rather:
\begin{itemize}
\item{+:} The $J_R^+$ components of 
actually both even and odd objects are 
fixed to $\pm \frac{a \alpha'}{2}$ as 
a remaining gauge choice after the 
conformal gauge has been used to gauge fix 
to the largest extend. This would have been the 
infinite momentum frame choice basically,
once we assumed that the distances in say 
$\sigma$-variable per object were (put) 
equal for all the objects. It really 
means that number of ``objects'' represent
the $P^+$-momentum of the string 
associated with those objects.

\item{$-$:} Next the components $J_R^-$ are 
fixed from the requirement gotten from 
the constraints in string theory, namely 
that $(J_R^{\mu})^2 =0$. This condition
fixes the $-$component (essentially energy)
in terms of the 24 transverse components 
$J_R^i$ and the gauge fixed $J_R^+$. 
Remembering that the ``odd objects'' are
constructed from the even ones by means of (\ref{oddfromeven}) we can write the $-$components as :
\begin{eqnarray}
&& \hspace{-4cm}
\hbox{For even objects:}\\
J_R^-( even \; I) &=& 
\frac{\sum_{i=1,2,...,24}(J_R^i)^2}{2 \cdot a\alpha'/2}\\
&& \hspace{-4cm} \hbox{For odd $I$ object(constructed):}\\
J_R^-(odd \; I )&=&\frac{\pi^2\alpha'\sum_{i=1,2,...,24}
(\Pi_R^i(I+1) - \Pi_R^i(I-1))^2}{a}
\end{eqnarray}
  \end{itemize}

It may be interesting to have in mind that
from the point of view of our Hilbert 
space description with a Fock space only 
having ``even objects'', and even those 
only with their transverse - the 24 
components - the odd objects as well as 
both the + and the - components are just
``mathematical constructions'' simply put 
up as mathematicians definitions. In this manner
the two of the 26 dimensions are pure 
``construction''! as well as half the 
number of objects. 

It were basically by means of these 
``constructions'' for a cyclical chain of
first even, then filled out by odd ones 
in between, that we in our previous 
article\cite{Novel} checked the spectrum
of masses. We ran, however, into a slight
species doubler problem: Because of our 
discretization of the 
$\tau_R$-variable we were seeking a 
spectrum of latticized theory 
(in one spatial dimension, the $\tau_R$),
and thus we got according to our theorem 
that one gets species doublers when 
seeking to make only right mover in fact
a species doubler\cite{19}. In order to 
get rid of that we propose to impose 
a continuity rule as a postulate.

\subsection{The Non-Parity 
Invariant Continuity Rule}   
\label{s:continuity}
The continuity rule which we saw earlier we had 
to impose to avoid a doubling of the 
usual string spectrum in our model is 
actually just the continuity rule, which
you would any expect. Crudely it just is
that you require the variation of the 
object $J_R^{\mu}$ or $J_R^i$ to vary 
slowly from object to the next object.
So physically it is extremely reasonable 
to assume this continuity rule. But we 
assume it - and have to assume it so - 
for both {\em even} and {\em odd} 
``objects'', and then because of the 
antisymmetry of the definition of the 
odd $J_R^i$ in terms of the conjugate 
of the even ones, we obtain a condition 
that {\em is not symmetric under the 
shift of sign of the object enumeration 
number $I$}. Intuitively you expect that
if a chain of numbers $J_R^i$ say, 
enumerated by $I$ vary smoothly with $I$
counted in positive direction, then it 
should also vary smoothly, if we count in the opposite direction. Because of our 
`strange'' definition of the odd object 
$J_R$-values, however, the continuity 
concept we are driven towards does {\em
not} have this intuitive property of being
inversion invariant. Let us in fact 
write our smooth variation or continuity 
requirement for three successive 
``objects'' in the chain - with an odd 
one in the middle say -
\begin{eqnarray}
J_R^i(I+1)& \approx& -\pi\alpha'(
 \Pi_R^i(I+1) - \Pi_R^i(I-1) ) \approx
J_R^i(I-1)\label{continuity} 
\end{eqnarray}
  
Imposing this non-reflection invariant 
continuity rule not only is a way to at
least assume away the species doubler from
the lattice, but it also gives an 
orientation to the $\tau_R$-variable.
For instance when we below shall match 
wave functions for strings in initial and
final states to calculate the overlap, 
this oriented continuity condition can 
let us ignore possible overlaps, if the 
two, to be matched, chains of ``objects''
are not oriented - in terms of the 
continuity condition - in the matching 
way. This rule reduces significantly the
possibilities for forming overlap 
contributions. From a symmetry point of 
view it may be quite natural that working
with only right mover say there should be
some asymmetry under reflection.

Thinking, however, on our model as the 
fundamental theory representing a seeming 
world with a string theory, it means that 
this rather strange ``continuity 
principle'' not being reflection invariant
has somehow to be imposed by the laws of 
nature. But now as already stated the 
Hilbert space structure and the dynamics 
in terms of ``even objects'' are just
the free massless scalar theory, and 
there is no place for such a reflection 
non-invariant continuity condition, except
in initial and ``final state conditions''
So in terms of our ``even objects''
we must have a truly rather funny initial 
state assumption: The ``even objects''
sit in chains that are continuous or 
smooth in our special sense in one 
direction, but therefore cannot be it 
in the opposite direction!

Of course in some way this continuity is
a description of the continuity of the 
strings, their hanging together.
   
     \section{Sketch of Calculation}
As one - and perhaps the most important -
tests of whether 
our string field theory in fact leads
to the Veneziano model scattering amplitude
(at least up to some overall factor, which
we shall leave for later works, and modulo
a rather short treatment only of the 
rather important appearance of the 
Weyl anomaly in 2 dimensions, which 
happens to be where the dimension of 
26 is needed in our calculation).
We shall also reduce the troubles of 
calculation by choosing a very special
Lorentz frame, something that would not
in principle have mattered provided the 
theory of ours had been known to be 
Lorentz invariant. However, since we
use infinite momentum frame - which is
not manifestly Lorentz invariant - it is
in principle dangerous to choose a 
special frame. 
    
\subsection{The Veneziano Model to 
be Derived}
Let us shortly - and especially with also 
a purpose of the extra factor in the 
integrand, for which we shall need the 
anomaly for the Weyl symmetry - recall 
what Veneziano model amplitude we shall 
derive, if we shall claim that it is a 
success for supporting that our model/our 
string field theory is indeed describing 
string theory of the bosonic 25+1 
dimensional type, the most simple string 
theory having though as a little problem, 
a tachyon. Since it is the simplest and 
historically the first to have a 
Veneziano amplitude for four external 
particles
\cite{21}, firstly later we generalized 
to larger number of external particles
\cite{12}, we 
shall start by deriving the 
Veneziano model for four external 
particles, although not in the 
phenomenologically supports case of 
Veneziano, 3$\pi$ + $\omega$. Rather we 
consider here just four external tachyons
each having mass square 
\begin{equation}
m^2 = - \frac{1}{\alpha'}
\end{equation}     
where $\alpha'$ is slope of the - before 
inclusion of loops - assumed ``linear 
Regge trajectories'', the leading one of 
which has the expression 
\begin{equation}
\alpha(t)= \alpha(0) + \alpha'  t,
\end{equation} 
where 
\begin{equation}
\alpha(0) = -\alpha' m^2 =1. 
\end{equation}
The four point Veneziano model is 
basically given by the Euler Beta function,
which can be defined by the integral
\begin{eqnarray}
B(x,y) &=& \int_0^1z^{x-1}(1-z)^{y-1} dz\\
\hbox{being used say for}&&\\
B(-\alpha(t),- \alpha(s))&=& \int_0^1
z^{-\alpha(t)-1}(1-z)^{-\alpha(s)-1}dz.
\end{eqnarray}
In writing such four point amplitudes 
one uses normally the Mandelstam variables
\begin{eqnarray}
s &=& (p_1+p_2)^2 = (p_3+p_4)^2\\
t&=& (p_1-p_4)^2 = (p_2-p_3)^2\\
u&=& (p_1-p_3)^2 = (p_2-p_4)^2\\
&&\hbox{obeying the relation}\\
s+t+u&=& m_1^2 +m_2^2 +m_3^2 +m_4^2 =4m^2 = -4/\alpha'.
\end{eqnarray}
Here the four- or rather 
26-momenta $p_i = p_i^{\mu}$ $(i=1,2,3,4)$
are the 
external momenta for the tachyonic string
states we consider as scattering states in
the simplest case considered here, all 
four counted the physical way, i.e. with 
positive energies $p_i^0$ for the process
\begin{equation}
1+2 -> 3+4,
\end{equation}   
being considered.
Since we consider the case of pure strings
without any Chan-Paton factor giving 
 quarks at the ends, the full scattering
amplitude becomes a sum over three terms
of the betafunction form. In front there 
is a factor $g^2$ involving the coupling 
constant $g$ for the string or Veneziano
theory being its square $g^2$. We shall, 
however,
postpone the presumably a bit complicated 
but very interesting question of the 
overall normalization in our theory to a 
later article. Thus the full amplitude 
expected is
\begin{eqnarray}
A(s,t,u)& =
& g^2\left\{B(-\alpha(s),-\alpha(t))
\right.
\\
&&
\left.
+B(-\alpha(s), -\alpha(u)) + B(-\alpha(t),
-\alpha(u))\right\}\\
&=&g^2(\int_0^1z^{-2\alpha'p_1\cdot p_2
-4}(1-z)^{-\alpha'p_1\cdot(-p_4) -4}dz+\\
&&
\int_0^1z^{-2\alpha'p_1\cdot p_2
-4}(1-z)^{-\alpha'p_1\cdot(-p_3) -4}dz+\\
&&\int_0^1z^{-2\alpha'p_1\cdot(- p_4)
-4}(1-z)^{-\alpha'p_1\cdot(-p_3) -4}dz). 
 \end{eqnarray}  
Since we have chosen to set up our model
in what deserves to be called infinite
momentum frame and to use the gauge 
that each object carries the same $p^+$
or rather having the fixed value $J^+
= a\alpha'/2$ according to 
(\ref{gaugefixing}), our formalism is {\em
a 
priori highly non-Lorentz invariant}, and 
it almost requires a miracle for it to turn
out at the end Lorentz invariant. It is 
therefore non-trivial and a priori 
dangerous only as we have chosen in the 
beginning to 
compared our model to 
the Veneziano model in the special case 
that the four external particles have 
the same $p^+$ components,
\begin{eqnarray}
p_1^+ = p_2^+& =& p_3^+ = p_4^+\\
\hbox{and consequently}&&\\
N_1 = N_2 &=& N_3 =N_4,
\end{eqnarray}
where the (even) integers $N_i$ 
(i=1,2,3,4) denote the 
numbers of ``objects'' attached to the 
four external particles. This choice of 
a special coordinate frame leads to a 
 simplification of the term without poles
in the s-channel:
\begin{eqnarray}
g^2B(-\alpha(t), -\alpha(u))&=&\\
g^2B(-1 -\alpha'(p_1-p_4)^2, -1-\alpha'(p_1
-p_3)^2)&=&\\ g^2B(-1 +\alpha'
(\vec{p}_{T1}-\vec{p}_{T4})^2, -1 +\alpha'
(\vec{p}_{T1}-\vec{p}_{T3})^2)&=&\\
g^2\int_0^1z^{-2 +\alpha'
(\vec{p}_{T1}-\vec{p}_{T4})^2}(1-z)^{-2+
\alpha'(\vec{p}_{T1}-\vec{p}_{T3})^2}dz&&.
\label{intransverse}
\end{eqnarray}
Here we have denoted the ``transverse''
parts - meaning the first 24 components 
by 
\begin{equation}
\vec{p}_T = \{ p^i \}_{i = 1,2,...,24}.
\end{equation} 
The simplification comes about because the 
+- term in the contraction with the 
metric in, say, $(p_1-p_4)^2$ drops out
 because of our very special frame choice
so that $(p_1 -p_4)^+ =0$,and so the 
$(p_1-p_4)^-$ does not matter, and 
\begin{equation}
(p_1-p_4)^2 = - (\vec{p}_{T1}
-\vec{p}_{T4})^2.
\end{equation}
  
 \subsection{Amplitude in Our Model, 
Principle of No Interaction!}
Whereas in string theory there seems to be
an interaction between the strings, it is 
rather surprising - and a hallmark for 
our theory - that in the formulation of
ours in terms of the object states the 
S-matrix elements, that shall give the 
Veneziano amplitude as we shall show, is 
simply 
equal to the overlap! That is to say it
is calculated as if the genuine S-matrix
is just the unit operator. More precisely
the S-matrix $<1+2|S|3+4>$, that shall 
describe the 
scattering  of say, two incoming open 
strings 1+2 into two outgoing 3+4 is 
obtained by writing the states in our 
formalism - in terms of even ``objects''
- corresponding or representing 
the two string state 1+2, say $|1+2>_{eo}$
 and also to the 
two string state 3+4 corresponding 
state in even object space, say
$|3+4>_{eo}$,  
 and then simply 
one takes the overlap of these 
incoming and outgoing states:
\begin{eqnarray}
<1+2|S|3+4> &=& <1+2|_{eo}|3+4>_{eo}.
\end{eqnarray}      
Here the subindex ${}_{eo}$ stands for
``even objects'' and means the state 
described in our even object notation.
This means that in terms of our string 
field theory = ``even object formulation''
a scattering goes on without anything 
happening (whatever might happen 
in reality must have been thrown out 
in the construction of our string field 
theory model). Symbolically this 
formula for the S-matrix is shown on the 
figure 4:
\begin{center}
\includegraphics[clip]{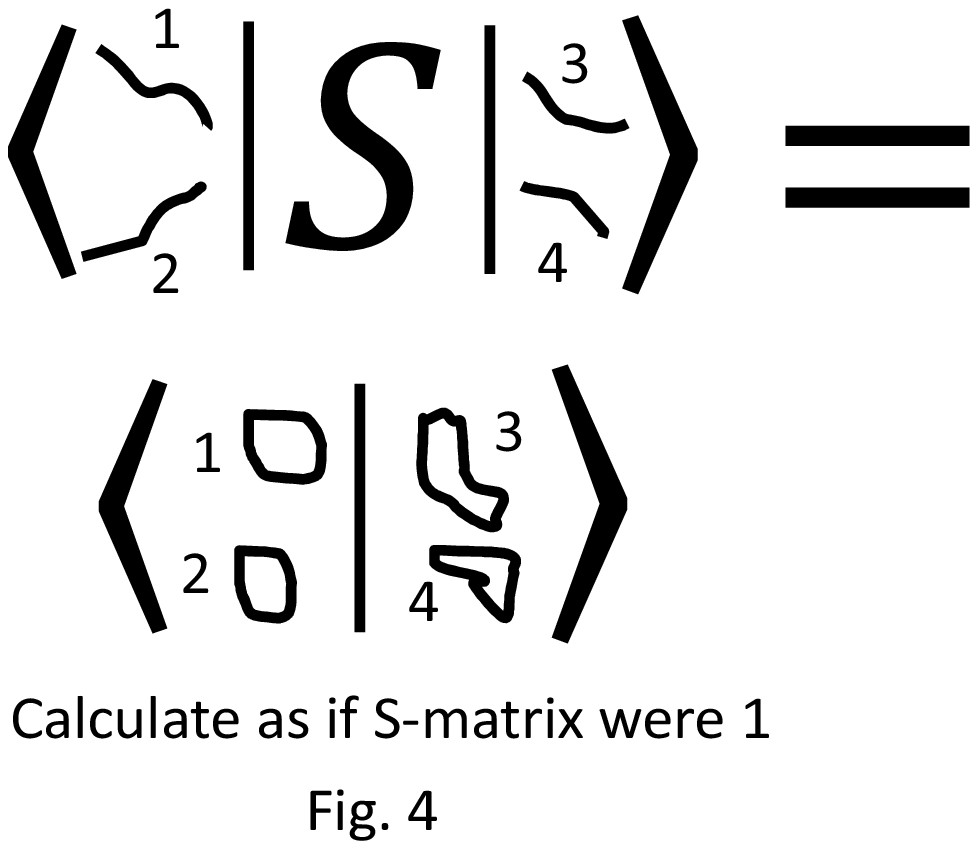}
\end{center}

\subsection{Procedure}     
The main tasks in order to evaluate the 
scattering amplitude are
\begin{itemize}
\item{A.} First we must evaluate in some 
useful notation the wave functions for 
the incoming and outgoing strings -
we shall in this article only consider 
scattering of two open strings coming in 
and two open strings coming-out, all
in the tachyonic ground states. 
\item{B.} We must figure out in how 
many different ways the ``even objects''
associated to the strings 1 and 2 in the 
initial state of the scattering can be 
{\em identified} with ``even objects'' 
associated with the final state strings.
\item{C.} For each way of identification
of every ``even object'' in the initial
state(i.e. associated with one of the 
incoming strings/particles) with an 
``even object'' in the final state 
(i.e. associated with one of these 
outgoing particles),
we have in principle two wave functions 
for the ``even objects'' and shall 
compute the overlap of these two wave 
functions.
\item{D.} Then we have to find the total
overlap by summing over all the different
ways of identifications, considered under 
B. 
\item{E.} This summation under D. will
turn out to be approximated by an integral
and we shall indeed see, that it becomes 
essentially the integration in the Euler
Beta function definition thus providing 
the Veneziano model.   
\end{itemize} 

In performing this procedure we make some
important approximations and 
simplifications:
\begin{itemize}
\item{a.} We shall assume that due to the 
continuity of the object series it is 
by far more profitable for obtaining a big 
overlap contribution to keep as many of 
the pairs of neighboring ``objects'' in 
the initial state, say strings 1 and 2,  
remain neighbors 
again in the final state. This means that 
we assume that the as contributions to 
the overlap dominating identification 
- in the sense of B. - are those in which 
the largest unbroken series of 
``even objects'' go from one initial state
string to one of the outgoing strings. 
This means the most connected or simplest
pattern of identification.

In fact the not yet quite confirmed though
speculation is, that the successively more 
and more broken up pattern of 
identification
of initial and final ``even objects'' will
turn out to correspond to higher and 
higher (unitarity correction) loops in 
dual models(=Veneziano models). Thus we 
expect, that considering only the least 
broken transfer of the ``even objects'' 
from the initial to the final strings 
shall give us the lowest order Veneziano
model (the original Veneziano model 
without unitarity corrections).
\item{b.} We shall of course use, that 
we take the limit $a \rightarrow 0$ and
correspondingly, that the numbers 
$N_1, N_2,N_3, N_4$ of 
``objects'' associated with  the various
strings go to infinity. Thus we can 
integrate over the number of objects in a 
chain going some definite way, say from 
string 1 to string 4.
\item{c.} To simplify our calculations we 
choose the very special case of the four 
strings - the two incoming 1 and 2 and 
the two outgoing 3 and 4 - all are 
associated with {\em same number of 
``even objects'' $N/2$} (and then of 
course  
$N$ `objects'' altogether). This 
assumption is with our letting the number
of ``objects'' be proportional to the 
$P^+$-component of the 26-momentum of 
the string in question, i.e. just the 
choice 
of Lorentz frame, so as to have all the 
four external strings/particles have the 
same $P^+$. So it looks like just being
a coordinate choice, but there is the 
little problem strictly speaking: that
our use of infinite momentum frame makes 
our theory not guaranteed to be Lorentz 
invariant. Anyway we do it only this 
non-invariant way in the present article
and leave it for  later, either to prove 
Lorentz invariance of our model, or to 
do it in a more general frame.     
 \item {d.} As a further strengthening of
point b. above about the chains coming 
in as long pieces as possible being 
dominant we remember, that our continuity 
condition (\ref{continuity}) was {\em not 
reflection invariant}. It would therefore 
be extremely little overlap, if we should
attempt to identify the ``even objects''
of a series in the initial state with a 
series in the final state in the opposite 
order. That is to say we require, that the 
orientation in the pieces of series 
going over as hanging together from 
initial to final state are kept. Otherwise
the contribution is assumed negligible.

Together b. and this item d. means that 
the dominant contributions come when 
possibly the longest connected pieces go 
over from one initial to one final without
changing orientation of the piece.  
\end{itemize}

We shall in the following seek a way 
to progress, that relatively quickly 
leads to string-theory-like expressions
and thinking. But the reader shall have 
in mind that even, if we shall approach 
string-theory-like expressions, we have 
at the outset
had a formulation - namely our string 
field theory - in which at first the 
stringyness is far from obvious. Rather 
it seemed that the stringy structure only
comes in with the initial and final states,
while the structure of our free massless 
scalar Hilbert space or Fock space is 
too trivial to contain any sign of 
being a string theory. It is therefore 
still interesting to calculate the results
of our theory, even if it quickly should 
go into to run along lines extremely 
similar to usual string theory.
  
\subsection{Construction of Wave 
Functions
for Cyclically Ordered Chains 
Corresponding to Strings} 
The wave functions for open strings were 
in fact investigated in our previous 
article \cite{Novel} in as far as the 
quantum system of $N$ objects forming 
a cyclically ordered chain corresponding 
to an open string were resolved into 
harmonic oscillators and thus a Gaussian 
wave 
function 
were obtained
in a high  
(of order $N$) dimensional space. The 
trick we shall use here  is to write the  
wave function of this character by means 
of a functional integral so reminiscent 
of the Feynman-Dirac-Wentzel functional
integral for a string propagation already
put into the conformal gauge, that we can 
say that we already managed to ``sneak 
in'' the string by this technology. 

In fact one considers in single string 
description functional integrals 
of the type:
\begin{eqnarray}
\int \exp(-\int_{A} (\vec{\partial}
 \phi(\sigma^1,
\sigma^2))^2 d\sigma^1 d\sigma^2) 
{\cal D}\phi,&& \label{functional}
\end{eqnarray} 
with some boundary conditions along the 
edge of the region $A$ say  in $(\sigma^1,
\sigma^2)$
space, over which the integral in the 
exponent is 
performed. We shall for our purpose of 
making an expression for the wave function 
in terms of our ``even objects'' for a 
string state consider that the region $A$ 
is taken to be a unit disk and at the 
edge we imagine putting a series
of ``even objects'' each being assigned 
a small 
interval along the circular boundary. 
Then we identify for example the object 
$J_R^i$ with the difference of the values 
of a $\phi^i$ taken at the two end points
of the little interval on the  
circle surrounding $A$ assigned to the 
object in question.
That is to say for say object number $I$
(here $I$ is even) having as its interval,
 say, the little 
region between the points on the circle 
marked by the angles 
\begin{eqnarray}
\theta_{beg}&=& 2\pi*\frac{I-1}{N}\\
\theta_{end}&=& 2\pi*\frac{I+1}{N}
\label{assinter}
\end{eqnarray}    
we identify (e.g.) the difference 
\begin{eqnarray}
J_R^i(I)& \stackrel{ident.}{=}& \phi^i
(\exp(i\theta_{end})) - \phi^i(\exp(i\theta_
{beg})),\label{difference}
\end{eqnarray}
where we have of course taken a new 
$\phi^i$ for each of the 24 $i$-marked 
components of $J_R$ and where we have 
identified the $(\sigma^1,\sigma^2)$- space with
the complex plane by considering $\phi^i$
a function of $\sigma^1+i\sigma^2$. 

The idea, which we seek to use here is 
that - possibly by some minor 
modifications, which we must state - we
should imagine, that we want to construct 
a prescription for obtaining a wave 
function
of the type $\Psi(J_R^i(0), J_R^i(2), ...,
J_R^i(N-2))$ as used in the expression 
(\ref{stateC_1}), describing say the 
ground state of a string in our formalism
by imposing a boundary condition 
- depending on a set of values for 
all the ``even objects'' in a chain - on 
the 
functional integral(s). With these 
boundary conditions imposed at the end
the functional integral become the 
wave function value $\Psi(J_R^i(0), 
J_R^i(2), ..., J_R^i(N-2))$ for the 
in the boundary condition used $J_R^i(i)$
values. 

Let us before fixing the details 
immediately reveal that we shall have an
extra boundary condition in the center 
of the disk $A$ at which point we shall
cut off an infinitesimally little disk 
and use the thereby opened boundary 
conditions to ``let in the (transverse
components of) the 26-momentum of the 
string in question''. This ``letting in'' 
means in principle that we put on the 
inside of the little circle a series of 
$J_R$ arranged to correspond to string 
with the right 26-momentum, but due to 
the smallness of the little circle the 
details except for this total momentum 
does not matter. In the figure we 
illustrate this situation on which we 
think: The line ending at the center and
ascribed a $P$ symbolize the just 
mentioned ``in-let'' in this center. The 
small
tags on the edge of the disk symbolize 
the attachments of the ``even objects'',
the values of which are used to fix the
boundary conditions for the functional
integral.            
\begin{center}
\includegraphics[clip]{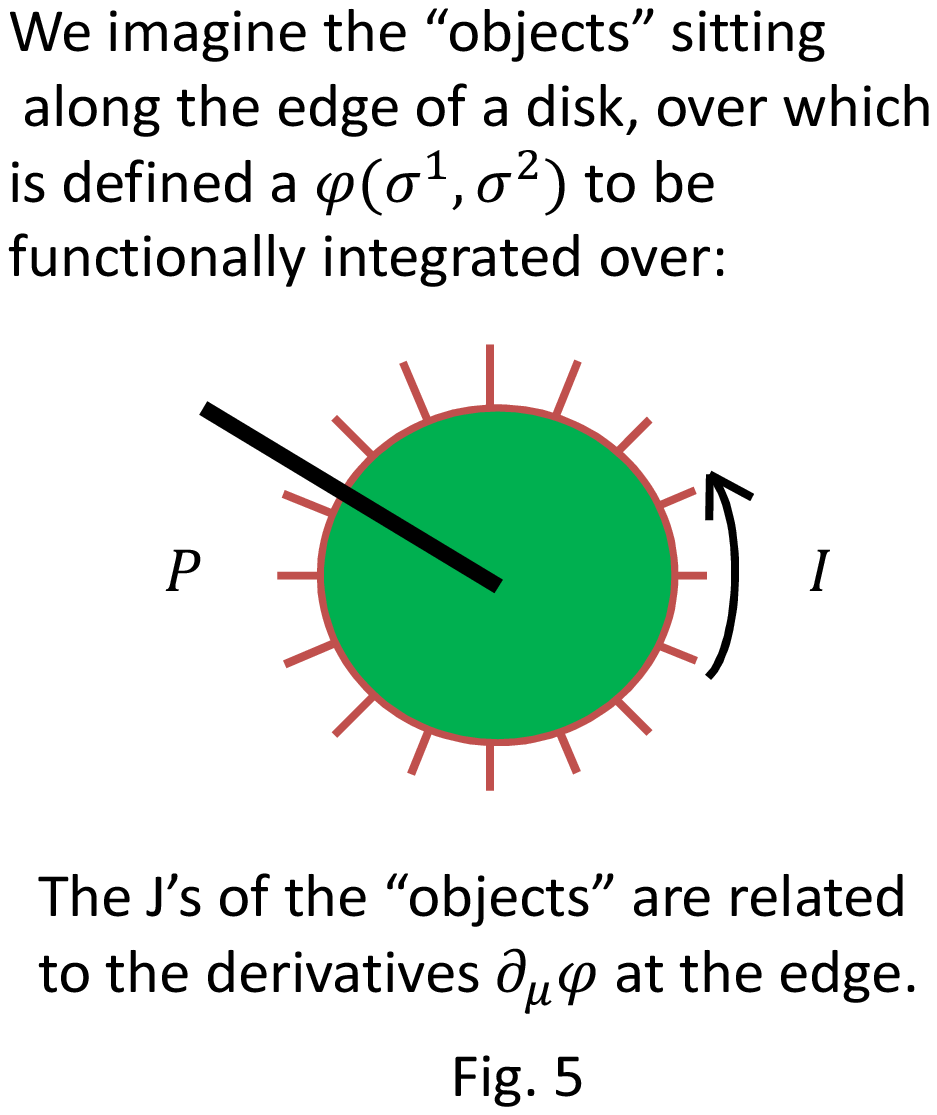}
\end{center}

Crudely the idea behind this procedure 
could be considered to be that {\em we
let a (here open) string propagate along 
during an imaginary time (say an imaginary
$\tau$), whereby only the lowest mass state
survives.} The heavier eigenstates of mass 
decay in amplitude faster that the 
lightest 
state by such an imaginary being spent.
Thus one gets after infinite imaginary 
``time'' the ground state selected out.
Thus investigating the wave function 
reached after such an infinite imaginary 
``time'' it should turn out being the 
ground state wave function, and so we 
should be able to use it as the 
$\Psi$ we want, if we want the wave 
function for a ground state string (the 
tachyon). Then the idea is of course to 
write the infinite imaginary ``time''
development by means of 
Wentzel-Dirac-Feynman path way integration.

Thus we get into our way of presenting the
wavefunction $\Psi(J_R^i(0), J_R^i(2),...,
J_R^i(N-2))$ a functional integral with at
first having a region, like $A$, being 
an infinite half cylinder. The axis along 
the half infinite cylinder is the 
imaginary part of the infinite imaginary
`time'', while the coordinate around the 
cylinder is rather the parameter, $\tau_R$,
enumerating the objects in the cyclically
ordered chain of objects associated with 
an open string. 

Then the type of functional integral 
here considered is ``essentially'' 
(meaning except for an anomaly becoming
very important at the end) invariant 
under conformal transformations of the 
region 
$A$. Thus ignoring - or seeing that they
are not there in the case considered - 
anomalies
we can transform the infinite half 
cylinder into the unit disk with the 
little hole in the middle, through which 
we ``let in'' the momentum of the string.  

Note how the string here comes in (only):
We got to a functional integral strongly 
related to what one usually work with 
in string theory, {\em just with the 
purpose of constructing a wave function
$\Psi(J_R^i(0), J_R^i(2),..., J_R^i(N-2))$
describing the string state in terms of 
our objects.} But there is nothing 
``stringy'' in our Hilbert space structure of our object-theory. The string only 
comes in via this wave function. 

But of course it still means, that after 
we have got this wave function in, we get
our calculations being so similar to
usual string theory, that we can almost 
stop our article there, and the string 
theorist may have exercised the rest so 
often, that we do not need to repeat.
But logically we have to repeat because 
we are logically doing something else:

There is in our formulation in terms of
the `objects'' a priori no strings. We 
are on the way to see, that after all the 
strings must be there, because otherwise 
it would be strange, that we just get the 
Veneziano amplitude for scattering.  
   
\subsection{Adjustment of the Details of the Functional Integral}
A few details about the functional 
integral
may be good or even rather important to 
have in mind:
\begin{itemize}
\item{I.} As long we - as here -  just 
seek to write  the exponential for the 
wave 
function (which as we know for harmonic 
oscillators have the Gaussian form -
of an exponential of a quadratic 
expression in the $J_R^i(I)$'s (even $I$) 
-) we could use the old proposal by David 
Fairlie and one of us (HBN)\cite{Fairlie} 
of evaluating 
the exponential as the heat production 
in a resistance constructed as the surface
region $A$ as a conducting sheet with 
specific resistance $\pi (2?)\alpha'$. 
Then 
one shall identify the boundary 
conditions by letting the current running
out at the interval assigned to a certain 
``even object'' be equal to the $J_R^i(I)$
for that ``even object''.
\item{II.} There a is little problem, 
which 
we have to solve one way or the other 
with getting the ``continuity condition'' 
(\ref{continuity}) discussed in 
\ref{s:continuity}. Having fixed only the 
boundary condition to the ``{\em even}
objects'' through their $J_R^i(I)$ but not
involving the conjugate variables 
$\Pi_R^i(I)$ there is of course no way 
in which the strange non-reflection 
symmetric continuity condition of our 
could be imposed. Concerning the classical
approximation one may actually find out 
that one easily can find the classical 
$\phi^i$ solution over the complex plane 
introduced above after the formula 
(\ref{difference}) which reflects the 
continuity condition as well as you can
require for a classical solution by {\em 
extracting only the analytical part of 
the saddle point for 
$\phi^i(\sigma^1+i\sigma^2)$}.

Indeed one might - and we probably ought
to do it - construct a model, in which we
use both even and odd $J_R^i$'s on the 
boundary, in the sense that we assign 
only half as long intervals on the 
border for 
each object - meaning that we replace 
(\ref{assinter}) by
\begin{eqnarray}
\theta_{beg}&=& 2\pi*\frac{I-1/2}{N}\\
\theta_{end}&=& 2\pi*\frac{I+1/2}{N}
\end{eqnarray}  
and use it for both even $I$ and odd $I$.

But now what are we to impose for the 
odd object intervals on the disk border?
We want to obtain a wave function $\Psi
(J_R^i(0), J_R^i(2), ...,J_R^i(N-2))$ 
expressed {\em only} as a function of 
the ``{\em 
even} object'' $J_R^i$'s, while  
{\em no $\Pi_R^i$ are accessible among the 
variables, on 
which the wave function depends}. 

However, in functional integrals one 
can easily extract what corresponds to the conjugate variable; they are so to speak
related to the time derivatives, by 
relations of the type that the conjugate 
to a variable $q$ in a general Lagrangian 
theory is given by 
\begin{equation}
p=\frac{\partial L}{\partial {\dot{q}}}.
\end{equation}  

On the other hand the continuity condition
tells us that approximately the odd 
$J_R^i$'s can be replaced by their even 
neighbors. Thus the proposal is 
being pointed out that we identify the 
appropriate time derivatives with the 
values of the neighboring even 
$J_R^i(I)$'s. Putting up this proposal 
is rather easily seen to correspond to, 
that the boundary condition relating 
$\phi^i$ near the boundary to the even 
object $J_R$'s, which  we are allowed to 
use,   
get decoupled from say the anti-analytic 
component in $\phi^i$. So with such 
a boundary inspired by the non-reflection 
invariant continuity condition would lead 
to an arbitrary solution for say the anti-analytical part, while the analytical part
would get coupled. We should like to 
develop this approach in further paper(s),
but it may not really be needed.

Instead of seeking to put our continuity 
condition (\ref{continuity}) into the 
functional integral formalism, we here 
shall use it is as a rule for which 
pieces of cyclically ordered chains can 
be identified, and then we shall get only 
oriented two dimensional surfaces - 
looking formally like string-surfaces
for closed oriented strings although 
what we are talking about are {\em open 
strings}
 (but remember that we get the diagrams 
for 
open look like the ones say Mandelstam 
have for closed) -.   

\item{III.}  Although it is in fact 
functional integrals like  
(\ref{functional}), that we basically need,
it is so that such a functional integral 
has divergences. These divergences must 
in principle be cut off. But now it turns 
out that the cut off necessarily comes to 
depend on a metric. Therefore we should 
rather write our functional integral
(\ref{functional}) as if depending on a 
metric tensor $g_{\alpha\beta}(\sigma^1,
\sigma^2)$ in
 the 2-dimensional space 
time, although it formally would look that
there {\em is actually no such dependence
on the metric, at least as long as we 
just scale it up or down by Weyl 
transformations}. This seemingly metric
dependent functional integral looks like
\begin{eqnarray}
\int \exp(\int g^{\alpha\beta}
\partial_{\alpha}\phi(\sigma^1,\sigma^2)
\sqrt{g}d\sigma^1d\sigma^2) {\cal D}\phi,
\end{eqnarray}  
where then boundary conditions and region 
of the 
$(\sigma^1,\sigma^2)$-parameterization must
be further specified. The cut off 
procedure should also be specified; it
could for instance be a lattice curt off,
a lattice in the $(\sigma^1,\sigma^2)$ 
variables, say. Then the importance 
of the metric is that you need the metric 
to describe the lattice spacing. Note 
though also 
that formally a 
scaling of the metric/Weyl transformation
\begin{equation}
g_{\alpha\beta} \rightarrow 
\exp {2\omega} g_{\alpha\beta}, 
\end{equation}  
even when the scaling function 
$\exp{2\omega}$ 
depends on the $(\sigma1,\sigma^2)$ does 
{\em seemingly} not change anything, 
because the 
determinant $g$ of the two by two matrix
$g_{\alpha\beta}$ just scales with 
$\exp{4\omega}$
so that the square root just compensates 
for the scaling of the upper index 
$g^{\alpha\beta}$ metric. The Weyl 
transformation symmetry is {\em only } 
broken by the cut off (the lattice) 
depending on $g_{\alpha\beta}$. It is via 
this cut off the anomaly can come in.
        \end{itemize}

\subsection{Overlap Contributions}
The crucial step in calculating the 
Veneziano model amplitude in our 
model/string field theory is to see what 
are the 
possibilities for identifying all the 
even objects associated with the initial
strings/particles 1 and 2 to the ones 
associated with the final state 
strings/particles 3 and 4 in a way that 
to the largest extend keep neighboring 
(or better next to neighboring, since we 
only consider the ``{\em even} objects'')
``even objects'' going into neighboring 
ones in the same order(same succession).

To simplify the possibilities, we have to 
consider what we have chosen to assume - 
basically by appropriate choice of 
coordinate system - namely  that each of 
the four 
strings or particles are associated with 
the same number of ``objects''. We may 
remember that by our gauge choice the 
number of ``objects'' $N$ associated with
say an open string is proportional to 
the $P^+$ component of its momentum, so 
that choosing a frame, wherein all the 
four 
external particles have equal $P^+$ 
implies that they have an equal number of 
associated ``objects'' also.

Now to keep the ``objects'' most in the 
succession they already have in the 
initial state also in the final state we 
must let connected pieces of 
"even objects'' pass from say string 1 to 
string 4. Then the rest of the 
``even objects'' associated with string 1
must go to string 3. Now the 
``even object'' numbers on string 2 that 
must go to respectively to 3 and to 4 is
already fixed for what happened for 
string 1. Since they have to sit in 
succession and a cyclic rotation of the 
cyclically ordered chains is the very 
last rudiment of gauge choice, there is no 
physically significant freedom in the 
identification except for the starting 
number of how many objects go from 1 to 4.

On the figure it is illustrated how 
different series of ``even objects'' from 
1 or 2 
marked with some signature are refound 
- with
same signature - in 3 or 4. The idea of 
course is that each of the four series 
marked by the four different signatures 
are refound in both initial (1+2) and final
(3+4) states, and really are the same.
It is understood that the series of 
``even objects'' identified to be in 
both initial 
and final states are identified 
``even object'' for ``even object'' in 
same 
succession.  

\begin{center}
 \includegraphics[clip]{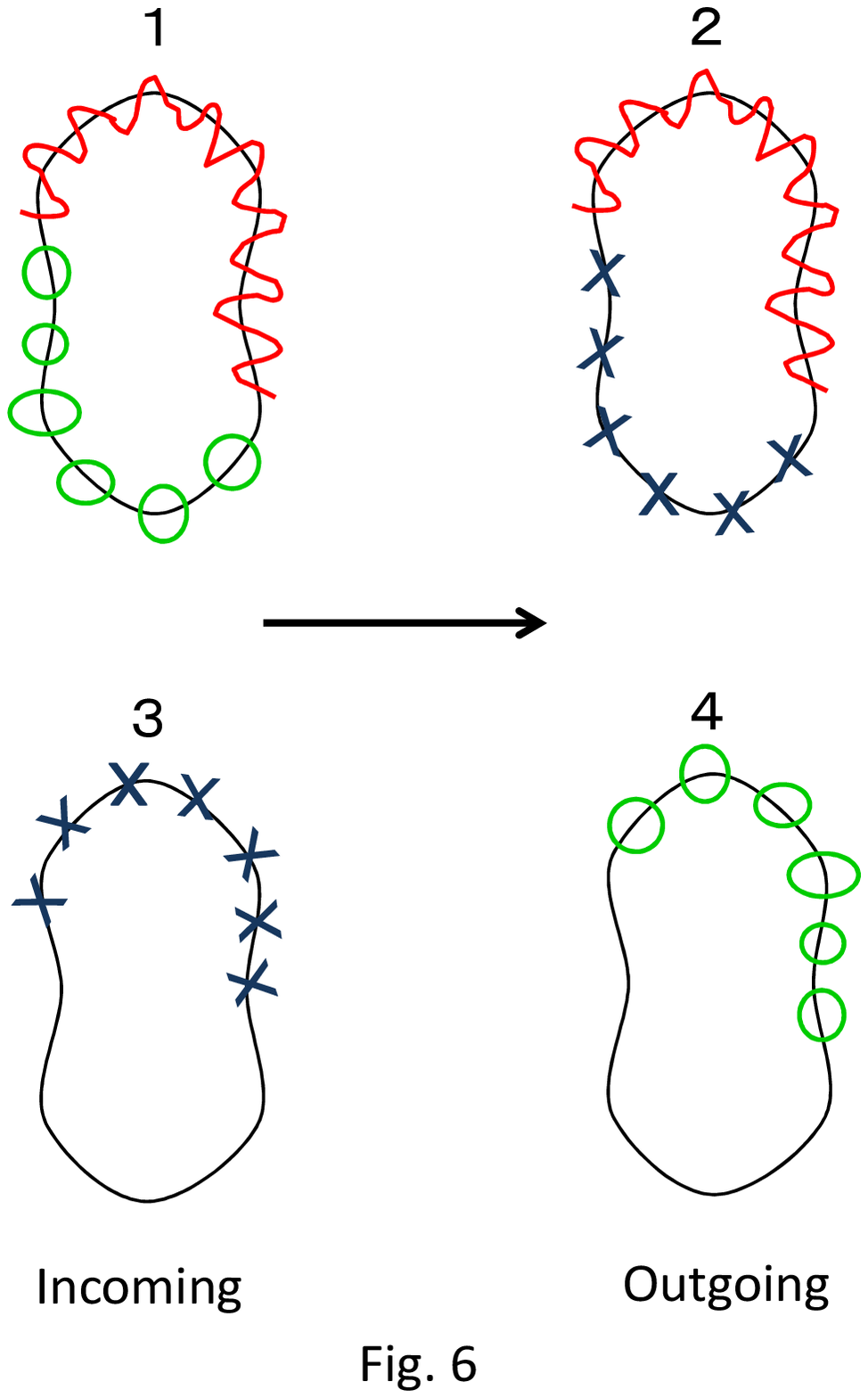}
\end{center}

To get the contributions to the overlap 
- and thereby amplitude - from all the 
physically different ``identification
ways'' one shall sum over the various 
values, a non-negative integer  
number, of ``even objects'' from 1 that 
are refound in 4. Since such numbers are 
of order $N$ - which means it goes to 
infinity as our cut off parameter $a
\rightarrow 0$ - the actual overlap 
contribution from each separate value of 
the number summed over varies slowly and
smoothly (we may check by our calculation)
and we can replace it by an integral over 
say the fraction of the ``even objects''
in 1 (i.e. associated with 1) that are 
identified with `even objects'' 
associated with 4. It is this integration 
that shall turn out to be the integration 
in the integration in the Euler Beta 
function making up the Veneziano model.

\subsection{The Overlap for One 
Identification}
But before integrating or summing we have 
to write down the overlap as obtained, if
one only includes the possibility of one
single ``identification'' (correspondence
between the ``even objects'' in the 
initial state 1+2 with them in the final 
3+4). This overlap of two states 1+2 and 
of 3+4 with a fixed ``identification'' is
of course simply the Hilbert product
of the two states of the set of 
$N/2 +N/2$ ``even objects'' - at least 
if one ignores
the low probability of two 
``even objects'' in say 1 and 2 being in 
the same state - so that we calculate it
as an inner product in an $N/2+N/2$ 
particle/``even object'' system. It 
becomes an inner product of the form
\begin{eqnarray}
&&
\int \Psi_{3+4, with identification I}^*(
(J_R^i(0), J_R^i(2),..., J_R^i(N-2))|_1,
 (J_R^i(0), J_R^i(2),..., J_R^i(N-2))|_2)
\nonumber  
\\
&&
\times 
 \Psi_{1+2}(
(J_R^i(0), J_R^i(2),..., J_R^i(N-2))|_1,
 (J_R^i(0), J_R^i(2),..., J_R^i(N-2))|_2)
 \nonumber  
\\
&&
\times 
\Pi_{i,I=0,2,...,N-2, k=1,2}dJ_R^i(I)|_k.
\label{inner}  
\end{eqnarray}       

Now the crucial point of our technique is 
that this inner product integration over 
the $J_R^i$-values for all the ``even 
objects'' associated with 1 or 2 (and 
identified with ``even objects'' in 3 
or 4) when the wave functions are written
as the functional integrals, we use, can 
be interpreted as just gluing functional 
integral regions together. The point is 
that the functional integrals basically 
are just - when cut off - integrals over 
$\phi^i$ values in all the different 
``lattice'' points along  the region $A$ 
boarder say.
At the boarders there is specified linear
relations of the $\phi^i$ values there 
- or rather the derivatives, but they are 
also linear relations - to the $J_R^i(i)$'s
assigned places on this border. One now 
has to argue that apart from an overall
constant factor we can consider the 
integration over the $J_R^i(I)$'s in 
(\ref{inner}) going in as part of the 
functional integration in a functional 
integral in which the regions $A$ for the 
two sides (initial and final) are glued 
together to one big functional integral.
Since the integration over the 
``even object''-variables $J_R^i(I)|_k$
have now been interpreted as part of the 
functional integration, the new resulting 
functional integral has no longer any 
boundary conditions associated with such
$J_R^i(I)|_k$'s. Rather the `big' 
functional resulting - and expressing the 
overlap for a specific ``identification'' -
only has as boundary conditions the inlets
of the external 26-momenta(or rather
their transverse components only),$ P_1,
P_2, P_3, P_4$. 

One should notice how this picturing by 
functional integrals come to look really 
extremely analogous to gluing together 
strings. There is though a little
deviation from the usual open string 
theory at first, because we have 
cyclically ordered chains  
topologically 
of form as 
circles as would be closed strings to 
represent the open strings. Corresponding
to this little deviation we get at first 
that final contribution from 
``identification'' of ``even objects''
 between final and initial states, becomes 
conformally equivalent to a Riemann sphere
with the four inlets from the four 
external strings being attached to this 
Rieman sphere. This is what you would 
expect for closed string scattering in 
the usual string theory, but {\em we} 
obtain
this for {\em open} strings! It turns,
however, out that all our four ``inlets''
- where the momentum boundary conditions 
are imposed - come by a calculation we 
shall sketch - to sit on a circle on the
Riemann sphere. Thus there is 
``reflection'' symmetry between the two 
sides of this sphere and mathematically 
our overlap for the fixed identification
come to be equal to a functional integral
as usually used for open strings. In this 
way our model has the possibility of 
agreeing exactly with usual string theory.

\subsection{Seeing the Hope}
A bit of imagination of how our 
topologically infinite half cylinders 
can glue together would reveal, that we
could arrange to get them pressed down in 
a plane but  
with - we must  stress though - in two 
layers.
In such a form we could have arranged that 
the result would look like a {\em double}
layer four string bands meeting along 
intervals with their neighbors but only 
in one point with their opposite string.
In order that we could bring it to look 
like this, we should put the two 
incoming strings opposite and the two 
outgoing strings opposite to each other. 
This would be the usual string gluing
picture for the open strings - 
just doubled, but that essentially does not matter - for the $B(-\alpha(t), 
-\alpha(u))$ term. This means that it is
extremely promising that we should 
obtain this term of the Veneziano model.

But !: 
\begin{itemize}
\item{1.} What happens to the other two terms
$B(-\alpha(s), -\alpha(t))$ and 
$B(-\alpha(s), -\alpha(u))$, which we 
should also have gotten, to get the full
Veneziano model?
\item{2.} We have in principle to check
that our model predicts the correct weight
factor on the integrand in the Veneziano 
model. We mean that the integrand, which 
we obtain does not only have the right 
dependence on the external momenta, but
also the right dependence as a function 
of the integration variable - which 
in our model comes from the summation over 
the different ``identifications''. 

Since in our model this integration comes
from the simple summation over 
``identifications'' our model has a very clear rule for what weighting to obtain 
and one just has to calculate carefully
not remembering the anomaly in the 
functional integral evaluation etc.

\item{3.} So far we were sloppy about 
the + and - components, or rather we only 
started calculating the factors in the 
integrand coming from the transverse 
momenta or transverse $J_R^i$ components 
so far.
\end{itemize}  

\subsection{Integrand weighting 
Calculation}   
If we want to evaluate the integrand 
meaning the contribution from one 
specific identification more carefully,
we have to be specific about how we for 
such an ``identification'' make the 
construction of the full surface on 
which the functional integration $\phi^i$
at the end gets defined. We obtain - using
the idea that the overlap integration 
can be absorbed into the functional 
integration - that we must glue together
four (either infinite half cylinders or)
disks corresponding to the four external
strings. To be concrete it is easiest to
represent the two final state strings 
3 and 4 by 
{\em exteriors} of a disk rather than 
by a disk as we represent the initial 
strings 1 and 2. The inlets for the 
final state strings are then at infinity
of the Riemann surface, while those of the
initial strings are at zero. 

Now however, we have two incoming and 
two outgoing, and so we are forced to 
work with a complex plane with {\em two} 
layers.

We take say one layer where the 
complement of the unit disk is put to 
be the essential disk for string 3, while
the other layer belongs then to string 4.
Similarly we must have two layers for the
initial strings, but now the important
point is that the initial and the final 
ones are to be glued together in a 
slightly complicated way, depending also
on the integration variable, which is 
essentially given by the number of ``even
objects'' going from 1 to 4 say.

Having settled on giving 3 and 4 each their
layer in the complex plane in the 
complement of the unit disk, we have 
let these outside unit disk layers be 
glued to the inside the unit disk ones 
associated with the two incoming 
particles 1 and 2 along the unit circle
of course. But now the length measured 
say in angle - or in number of 
``even objects'' proportional to the 
angle - along which say the layer in the 
inside assigned to string 1 has to be 
glued together with complement of disk
region layer assigned to string 4 along 
a piece of circle proportional to the 
amount/number of ``even objects'' passing
from string 1 to string 4 in the ``identification'' we consider. along the rest of 
the unit circle then of course the disk 
assigned to string 1 is identified with 
the outside disk layer connected to 
string 3. Similarly along the first piece 
of circle - where 1 is connected to 4 
- of course the layer of string 2 shall
be connected along the circle to the layer 
of string 3 (in the outside). 
Correspondingly along the ``rest'' of the 
unit circle the layer assigned to string 2
(inside) is attached (identified with) the
layer of string 4. In the figure you may see an attempt to give an idea of what to do
before having put on the final state 
strings associated with the complements 
of the unit disk in their two layers. But on the figure the inner layers are 
prepared for the gluing together.  
\begin{center}
\includegraphics[clip]{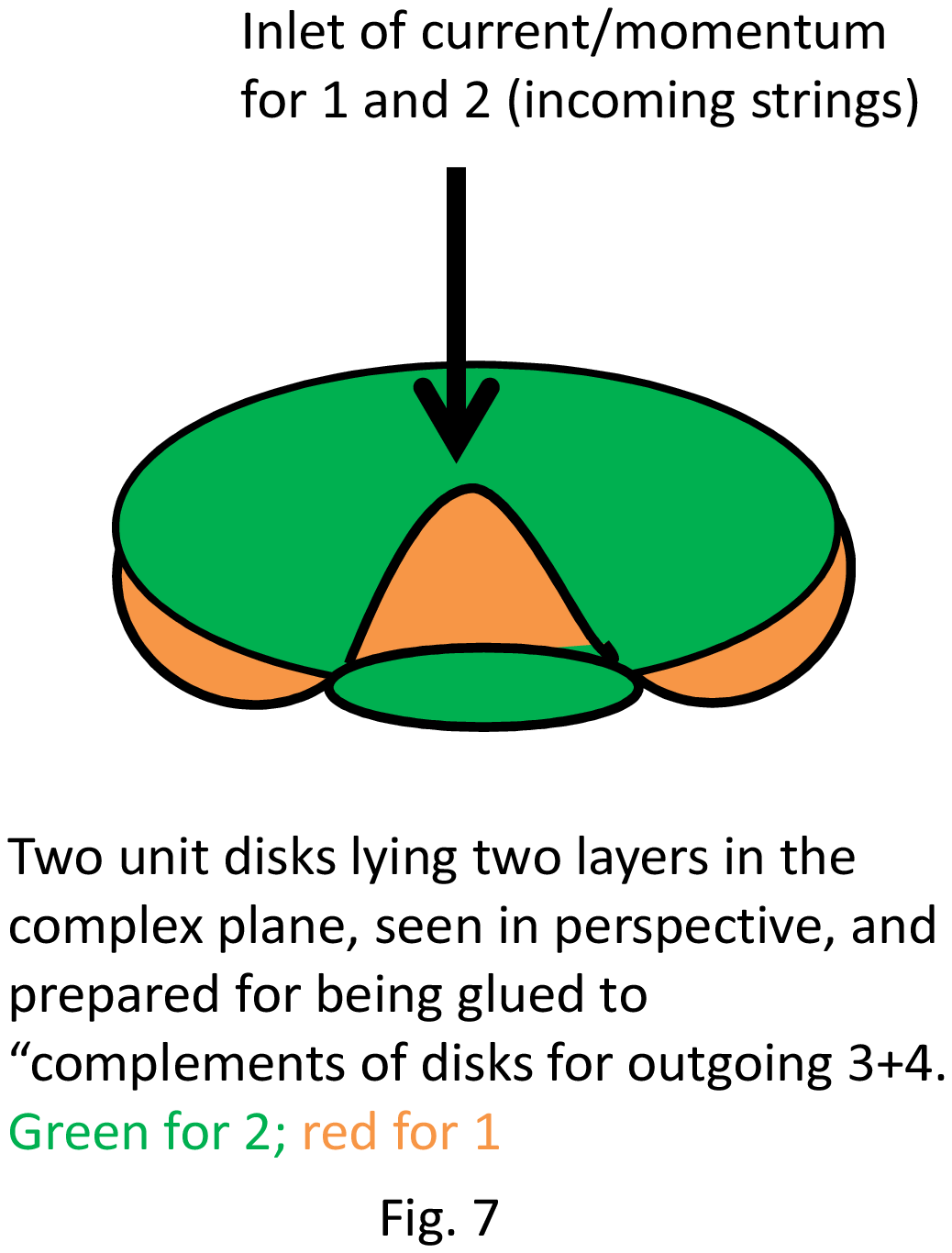}
\end{center} 

Now modulo the anomaly - i.e. naively - 
the functional integrals considered are 
conformally invariant, so that we are 
postponing the anomaly allowed to perform 
a conformal transformation of the 
combined region - now lying in two 
layers - of the four disks or complements 
of disks associated with the four external
strings, and the result of the functional 
integral being proportional to the 
overlap contribution from the 
``identification'' in question should 
not be changed. 

Since the angle $\theta$ (circle piece length) along which say layer 
of string 1 is identified with the layer 
of string 4 is proportional to the 
{\em number} of ``even objects'' we shall
simply integrate to get the an expression 
proportional to the full overlap (and thus to the Veneziano amplitude hopefully) 
integrate simply with the  measure 
$d\theta$.

At first it looks that we have a little 
problem by only having wave functions as 
functions of the 24 transverse coordinates
so that seemingly the + and - components
of the 26-momenta cannot appear in our 
hoped for Veneziano model integral. 
However, luckily for the term we actually 
obtain $B(-\alpha(u), \alpha(t))$ we found 
above in equation (\ref{intransverse}) 
that in fact all the terms in the 
exponents for $z$ and $(1-z)$ that depend
on the external 26-momenta could be 
arranged {\em to come only from the 
transverse momenta} provided we have 
made the very special frame choice that
the four external particles have the same 
$P^+$ components. So in this our 
simplifying the contributions from the + 
and - components turn out not needed. 
The point really was that just having the
gauge choice and the frame choice 
arranging the + component for the 
$p_1 -p_4$ and for the $p_1 - p_3$ become
zero the character of the $u$ and $t$
of having a + component multiplied with a 
- one made it enough to ignore but the 
transverse contributions.     

\subsection{The Conformal 
Transformation}
The way to evaluate the contribution to 
the overlap of $|1+2>$ with $|3+4>$ from
one ``identification'' is to rewrite it 
into a functional integral the region 
of which is composed from the four 
disks or 
disk complements corresponding to the four
external particles/strings. We obtain 
at first a manifold described as double 
layered in the Riemann sphere. It has two
branch points on the unit circle 
corresponding to the points where the 
``even objects'' on say 1 shifts from 
going to 3 to going to 4 (or opposite).
Basically we choose to map this doubled 
layered region by a map with two square 
root singularities at the two branch 
points. 



\subsection{Anomaly}
\label{s:anomaly}
The anomaly that gives us an extra factor
mutiplying the contribution from a single
``identification'' is usually written 
formulated as the trace anomaly
\begin{equation}
<T_{\alpha}^{\alpha}>=
- 
\frac{1}{48\pi}\sqrt{g} *R,
\label{anomaly}\end{equation}
(in the notation of our article with
Habara \cite{Habara}, wherein 
$
\sqrt{g}R = -2\partial_{\alpha}\partial
^{\alpha} \omega$ for the metric tensor 
of the form  $g_{\alpha\beta} =
\exp(2\omega) \eta_{\alpha\beta}$,and)
where $R$ is the scalar curvature of 
the metrical space given by the metric 
tensor (in two dimensions enumerated by
$\alpha =1,2;$). Here the energy momentum 
tensor is denoted $T^{\alpha\beta}$ is 
indeed for 
the theory of the field(s) $\phi^i$
which  had 
been Weyl invariant as it formally looks
like, and so the trace 
$T_{\alpha}^{\alpha}=0$ 
would
be zero. The symbol $\#fields$ denotes
 the number of 
fields $\phi_i$; it would in the 26=25+1
theory be 24. But the formula ( \ref{anomaly})
were made for a right moving and a left moving 
complex fermion field. Now since a scalar like the $ \phi$
obeys a second order equation it is in fact both containing 
a right mover and a left mover, but since it is real/Hermitean 
it only counts half the fermions used in ( \ref{anomaly} ). 
Using that fermions and bosons in 2 dimensions count the same 
we thus have for $d-2$ real scalar fields the anomaly 
\begin{equation}
<T_{\alpha}^{\alpha}>=
- \#(scalar-fields)* \frac{1}{96\pi}\sqrt{g} *R,
\label{anomaly2}\end{equation}
This anomaly can be seen to come in 
by having in mind that we want to perform
a conformal transformation - in fact the 
one corresponding to the analytical 
function 
\begin{equation}
f(z) = \sqrt{\frac{z-\exp(i\delta)}{z-\exp
(-i\delta)}} - \label{conf}
\end{equation} 
(here we used the notation that the 
end points of the cut along the unit 
circle seperating where sheet 1 connects 
to 4 from where it connects to the sheet 
associated with particle 3 were arranged 
to be $\exp(i\delta)$ and 
$\exp(-i\delta)$.)
and then the anomaly gives rise to 
corrections to the ``naive'' result that 
the functional integral is invariant 
under a conformal transformation.
In fact we may first have in mind that 
we shall evaluate the functional integral
(\ref{functional}) 
with a lattice or other cut off only 
depending on the internal geometry so that
it only can give variations depending on 
the metric tensor $g_{\alpha\beta}$, which 
under a conformal transformation only 
changes its scale locally as under a 
Weyl transformation (\ref{Weyl}).
 So what we only need to calculate to
obtain the effect of the anomaly is 
how the overall factor on the 
metric tensor varies under the conformal
transformation, we shall use (\ref{conf}).
Such a scaling is given by the numerical
value of the derivative of the function
(here $f$) representing the conformal
transformation,
\begin{equation}
g_{\alpha\beta} \rightarrow \Omega 
g_{\alpha\beta}\hbox{  where then }
\Omega = |\frac{\partial f}{\partial z}|^2.
\end{equation}  
This is to be understood, that the metric 
tensor describing the complex plane metric
in the $f$-plane is 
$\exp{2\omega} g_{\alpha\beta}$ or $\Omega g_{\alpha\beta}$ when the 
metric 
induced from the $z$-plane usual metric 
is $g_{\alpha\beta}$. 

It is easy to see that scaling the metric
tensor (locally) with an infinitesimal
scaling factor 
$\exp{2\omega}$ with $\omega << 1$ leads to a correction
to the logarithm of the functional integral
by $\int \Delta\omega T_{\alpha}^{\alpha}
d^2\sigma$. Since the trace $T_{\alpha}
^{\alpha}$ of the energy momentum 
tensor is only non-zero according to
(\ref{anomaly}) where there is a non-zero 
curvature,
and our two layered surface lies mostly in 
the flat complex plane, we only get 
contributions  to this Weyl transformation
local change of scale from the two 
(singular) branche points $\exp(i\delta)$
and $\exp(-i\delta)$, where the curvature 
$R$ has delta-function contributions.

We can without any change in value of the 
functional integral make a formal 
reparametrization from say the double 
sheeted complex plane to the single 
layered one by means of $f = \sqrt{\frac{
z-\exp(i\delta)}{z-\exp(-i\delta)}}$
provided one then use after the 
transformation the {\em transformed metric 
tensor}. With a conformal transformation
the transformed metric inherited from the 
z-plane into the f-plane will only deviate
from the flat metric $\eta_{\alpha\beta}$
in the f-plane by a Weyl transformation.
We know that there only shall be 
curvature - of delta function type - 
at those points in the f-plane that 
are the images of the branch points 
$z=\exp(i\delta)$ and $z=\exp(-i\delta)$,
and so the (Weyl transformed) metric 
reflecting the metric space from the 
z-plane into the f-plane, $\exp(2\Omega)
\eta_{\alpha\beta}$, i.e. $R=0$ outside
these two points $f=0\hbox{and}\infty$.

This means that the $\Omega$ outside those
two points in the f-plane must be a 
harmonic function  
of $f$, meaning the real part of an 
anlytical function. This outside the 
two points harmonic function shall though
have {\em singularities at the two 
points } on the f-plane (or the 
corresponding Riemann sphere rather)
delivering the delta-function 
contributions,
\begin{equation}
R = 4\pi \delta(Re(f))\delta(Im(f))
\hbox{at f=0 say.},
\end{equation}   
 At the branch points, we have points with 
the property that going around one of 
them in the z-plane or system of sheets
one get a return angle $\theta$ being 
$2\pi$ more than after the mapping into
the f-plane (or Riemann space). Thus the
integral over the curvature delivering 
this extra amount of parallel transport 
extra shift angle should in an 
infinitesimal region around the image of 
a branch point - say the point $f=0$ -
be $2\pi$. So with a notation 
with the rule of such an (excess) angle
being given as 
\begin{equation}
\int_{area} R \sqrt{g} d^2\sigma = 2\theta
\end{equation}   
with $\theta$ the extra angle of rotation 
on return, 
\footnote{In our notation we have the 
rule that going around an area and thereby
obtaining for for a parallel transported 
vector on return a rotation by an
angle $\theta$, that the integral over
this area 
\begin{equation}
\int_{area} R \sqrt{g} d^2\sigma = 2\theta
\end{equation}     } there will in the 
metric 
inherited from the $z$-plane in the 
f-plane be delta function contribution
to the curvature scalar $R$ at the points 
corresponding to the branch points 
$z=\exp(i\delta)$ and $z=\exp(-i\delta)$ 
in the 
$z$-plane, and thus $f=0$ and $f=\infty$
in the f-plane. One can easily see that 
because there is just $2\pi$ extra angle 
to go around such a branch point in the 
sheeted $z$-plane the delta-function 
contribution becomes e.g. for the $f=0$
point 
\begin{equation}
R = 4\pi \delta^2(f). 
\label{deltacurvature}
\end{equation}

If $r$ is the distance to the image of the
branch point, say the $f=0$ point, so 
that $r = |f|$, the solution to 
$R= -2\partial_{\alpha}\partial^{\alpha}
\omega$ for this delta function $R$ is 
a logarithm of the form
\begin{equation}
\omega(r) = \ln(r/K).
\end{equation}  
( Here $K$ is some constant in the sense 
of not depending on $r$)
That implies that taken at the point 
$r=0$ the $\omega(0)$ is logarithmically
divergent so that the integral to which
the anomaly of the logarithm of the 
correction to the integrand is 
proportioanal 
becomes divergent. However, we have 
anyway given up caculating in this article
the overall normalization of the 
Veneziano model, we hope to derive. We 
shall therefore be satisfied with only 
calculating the contribution in $\omega$ 
{\em that varies with the angle $\delta$}
over which we (finally) integrate. Now 
the conformal transformation mapping the 
two-sheeted $z$-plane into the one-sheeted
$f$-plane is 
\begin{equation}
f(z) =\sqrt{\frac{z-\exp(i\delta)}{z-\exp
(-i\delta)}}, \label{f}
\end{equation}
and so its logarithmic derivative
\begin{equation} 
\frac{df}{fdz}=\frac{1}{2}(\frac{1}
{z-\exp(i\delta)} - \frac{1}{z-\exp(-i\delta)}),    
\end{equation}
and the derivative proper
\begin{equation}
\frac{df}{dz}= \frac{1}{2}*\sqrt{\frac
{z-\exp(i\delta)}{z-\exp(-i\delta)}}
 *(\frac{1}{z-\exp(i\delta)} -\frac{1}
{z-\exp(-i\delta)}).\label{derivative}
\end{equation}

We are interested in a hopefully finite
term in the change in going from the 
$z$-plane simple metric to the one in 
the $f$-plane, which is the part of 
\begin{equation}
\Omega_{\hbox{z to f}} = 
\ln(|\frac{df}{dz}|)\label{Omega}
\end{equation}  
depending on the ``integration variable''
$\delta$. This means that we make precise 
the cut off by saying that we must make 
the cut off by somehow smoothing out the 
branch point singularity in a fixed way 
in the {\em $z$-plane}. This means that 
we perform a regularization by putting 
into our transformation a fixed distance 
$\epsilon$ in the $z$-plane marking the 
distance of $z$ to one of the branch 
points. That is to say we consider a 
little circle say of points around the 
exact branch point counted in the 
$z$-plane
\begin{equation}
z_{\hbox{on little circle}} = \epsilon
\exp(i\chi) + \exp(i\delta)
\end{equation}
 (or analogously using $\exp(-i\delta)$
instead of $\exp(i\delta)$.)
On this little circle we find that the 
scaling - Weyl transformation - going 
from the $z$-plane to the $f$-plane 
 using (\ref{derivative}, \ref{Omega})
\begin{equation}
\Omega_{\hbox{z to f}} 
= \ln(|df/dz|_{circle})\approx 
 \ln(\frac{1}{2\sqrt{\epsilon}
\sqrt{2sin(\delta)}})
\end{equation}
for the $z\approx \exp(i\delta)$ case. 
The $\delta$-dependent part is 
of course $-\frac{1}{2 } 
*\ln(\sin(\delta))$. This is the 
$\delta$-dependent part of 
$\Omega_{\hbox{z to f}}$ which comes 
into the anomaly correction for the
logarithm of the full (product over the 
24 values of $i$ of the ) functional
integral, according to
(\ref{anomaly}) and (\ref{deltacurvature})
of course with a coefficient proportional
to the number of truly present dimensions
in the functional integral - which is 
only the transverse dimensions 24 -.

Thus the $\delta$ dependent part of the 
anomaly ends up being in the logarithm of 
the contribution to the overlap from 
one value of $\delta$ ( meaning one 
``identification''):
\begin{eqnarray}
\Delta_{anomaly}\ln{\hbox{integrand}}&
=&\hbox{$\delta$ independent }+\\
+ (d-2) *\frac{1}{96\pi}* 
\frac{1}{2}\ln{\sin{\delta}}*(4\pi + 4\pi)&&\\\nonumber
&=& \frac{d-2}{24}*\ln{\sin{\delta}}+... \label
{anresult}
\end{eqnarray}

Now we should remember that we have 
decided in this article to go for the 
form of the amplitude but have left for 
further studies the over all normalization
of the amplitude. This means that the 
terms in the logarithm of the integrand 
of the hopefully to appear Veneziano 
amplitude which do not depend on the 
integration variable $\delta$ (which is
proportional to the number of (even)
objects from string 1 that goes into
string 4) but only so on the cut off 
parameter $\epsilon$ are neglected. 

Now the conformal mapping (\ref{f}) brings
the inlet points for external momenta for
the four external particles into the 
positions sketched on the figure:
\begin{center}
\includegraphics{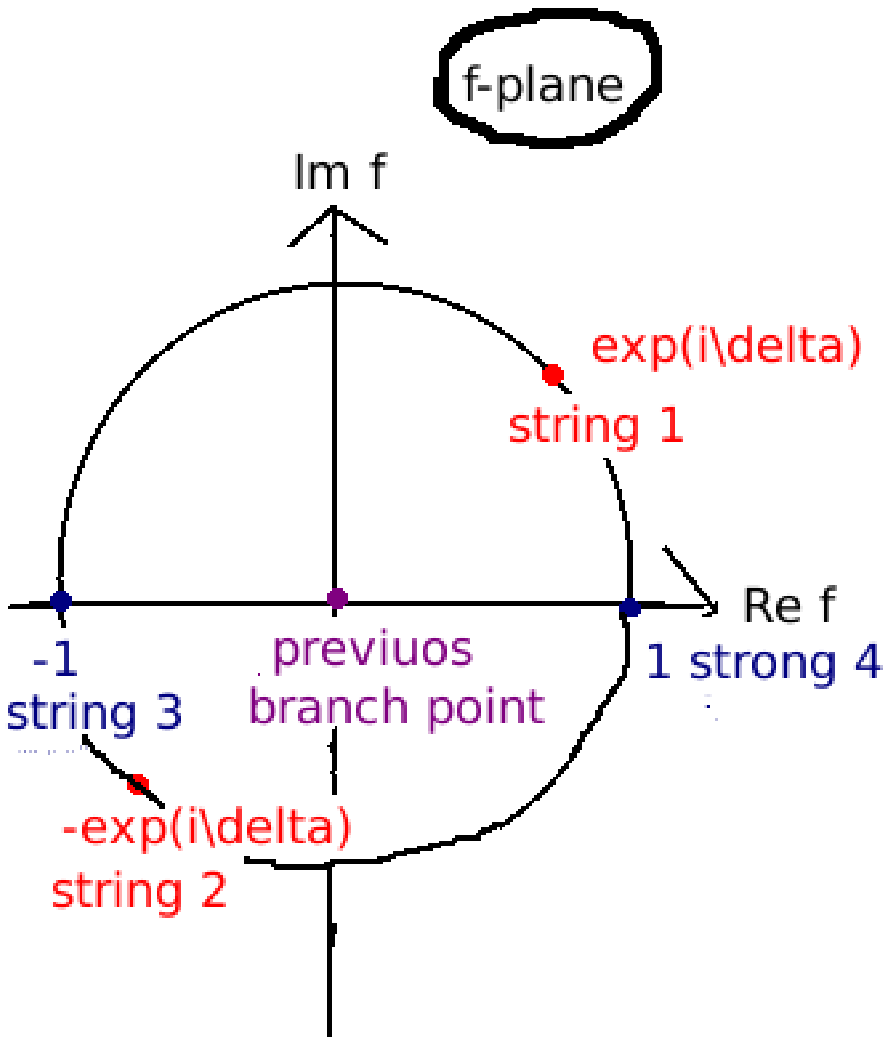}
\end{center}    

Imagining on this figure that one varies
the integration variable $\delta$, then 
the ``inlet'' points for the two 
incomming strings 1 and 2 will remain 
sitting opposite to each other one the 
unit circle and analogously the two final
state string inlet points 3 and 4. So 
the distances between 1 and 2 or between 
3 and 4 are constant as function of 
$\delta$ and so we can ignore the terms 
comming proportional to in fact $s=- (
p^i_1+p^i_2)^2+ ...= -(p^i_3+p^i_4)^2 +...$  from the heat 
production in the analogue model from the 
current running co as to depend on the 
distance between  1 and 2 or analogously 
between 3 and 4. In a similar way we are 
allowed with our decission to only keep
the $\delta$ dependent terms to ignore 
terms involving only one of the four 
inlet points. There are such contributions
but they depend on the inlet momentum 
squared (with a divergent coefficient),
but since only on one point the 
$\delta$-dependence is not there provided 
we cut off in a $\delta$ independent way
of course. The divergences form cut off 
of the anomaly correction connected with 
originallly - in z-plane - branch points
come only in to the extend that they get 
their cut off small circles scaled to a 
differnt degree depending on $\delta$.   

It is not difficult to see that to seek
identification of the $\delta$ dependent 
terms with the integration variable $z$ 
( not to be confused with our complex 
plane $z$ which is of course something 
different) in the Veneziano model we must
identify
\begin{eqnarray}
z &=& sin^2(\frac{1}{2}*\delta)\\
1-z&=& cos^2(\frac{1}{2}*\delta).
\end{eqnarray}

Very important for stressing how 
successful our model/rewritting is to
reproduce the Veneziano model integration 
measure in the $z$-integration correctly.
In our model this integration correponds 
to the summation over the discrete 
variable being the number of (even)
objects going from string 1 to string 4
and it is proportional to $\delta$, and
thus we at first simply the measure of 
integration is $d\delta$. But now to
compare with the usual Veneziano formula
expressions or our slight rewrittings of 
it (\ref{}) we must of course relate 
$d\delta$ to $dz$:
\begin{equation}
dz =   \cos(\frac{1}{2}\delta)\sin(
\frac{1}{2}\delta)d\delta 
\propto sin(\delta)d\delta 
\propto \sqrt{z(1-z)}d\delta.
\end{equation}   
   This happens to show that the 
correction factor from the anomaly 
needed to just compensate the factor 
comming from
\begin{equation}
d\delta = \frac{dz}{sin(\delta)}
\end{equation} 
would be just cancelled by the anomaly 
provided 
\begin{equation}
\frac{d-2}{24} =1 .
\end{equation}
This means 
that we get precisely the right Veneziano
model when the number of transverse dimensions $d-2 = 24$. The 
famous result that the bosonic string must exist in d=26 space time dimesions. 



\section{Our Shock; Only One 
Term}
When we went through the above sketched
calculation and arrived at {\em only} the 
one term proportional to $B(-\alpha(t),
-\alpha(u))$, which is the term without
poles in the s-channel, it were somewhat 
unexpected at first. After all we had made 
up a model essentially written out so as 
to {\em make it} the string theory and 
thereby the Veneziano model. Then it 
gave only one out of the three terms
it should have given. May be even more 
strangely, if we imagine investigating 
crossing symmetry it looks we would get 
a different term after what particles are incoming and which outgoing. So the term 
we got is not even properly crossing 
symmetry invariant. Nevertheless it were 
very encouraging that we got something so
reminiscent of the Veneziano model as 
simply one of the terms. 

We believe we have found a way to get 
the two missing terms also come out:

In fact we think that it is in a way the 
infinite momentum frame gauge, which we 
used, that is the reason for the 
surprising problem for our model: Really 
one may say that the infinite momentum 
frame is a method for avoiding having to 
think about the vacuum, which in quantum 
field theories is usually an enormously complicated state. In the infinite 
momentum frame type calculations you
imagine an approximation in which the 
particles have so high energy that they 
manage not to ``feel'' the vacuum. But 
such an approximation may not be a good 
one. So we thought it might be best 
somehow to introduce at least some 
rudimentary effects of a vacuum even 
though we want to continue to work with 
an infinite momentum frame formalism, 
especially an infinite momentum frame 
gauge/parameterization choice.

The idea, which we here propose, and 
which actually seems to help to obtain the
lacking two terms in the full Veneziano 
model amplitude, is to allow not only as
we did at first for ``objects'' with the 
+components $J_R^+=a\alpha'/2$ (a 
positive number), {\em but also allow 
``negative objects''} having rather 
their $J_R^+ = -a\alpha'/2$. At least 
with inclusion of such negative 
``objects'' you make it at least a 
possibility to have not totally trivial 
state with the property of the vacuum of 
having the ``longitudinal'' momentum 
$P^+=0$. The vacuum could so to speak 
consist of a compensating number of 
usual positive say ``even objects'' and 
corresponding number of ``negative even
objects''.

In fact it looks that we with such ``negative'' ``objects'' can imagine some of 
our strings represented by an ``extended''
cyclically ordered chain(replacement). 
Hereby we mean that it contains in the 
``extended'' cyclically ordered chain 
not only usual positive $J_R^+$ objects,
but also one or more series of negative 
$J_R^+$ objects, arranged so that the 
excess of positive ones over negative 
ones is proportional to the total 
$P^+$ component of the 26-momentum for
the string in question. With such 
``extended'' cyclically ordered chains 
representing some open string we obtain
the possibility of the negative part 
of say string 2 annihilating with part 
of the cyclically ordered chain of string
1. Similarly one of the final state 
strings could be produced with content in
its cyclically ordered chain of some 
series of negative objects having been 
produced together with some positive 
ones in another final state string.

By very similar procedure to the one 
used above to the term $B(-\alpha(u),
-\alpha(t))$, but now including the 
negative objects we seem to be able to 
produce the two missing terms.
The detail of the calculation to obtain
the full Veneziano amplitude/model will
appear soon by the authors \cite{22}.

\section{Conclusion and Outlook}
We have in the present article sketched 
how using our string field theory 
formalism in which strings are rewritten 
into be described by states of 
``even objects'' we can obtain the 
scattering amplitude to be the usual 
Veneziano model amplitude. It must though
be immediately admitted that we at first 
got {\em only one out of the three terms
expected}. However, introducing 
``objects'' that can have negative 
$J_R^+$-components and can function as 
a kind of holes for objects, we though 
believe, that it is promising to obtain
the whole Veneziano amplitude. Our model
or string field theory has previously been
shown \cite{Novel} to lead to the usual
mass square spectrum for strings. In 
this way we collect increasing evidence
that our formalism is indeed another 
representation of all of string theory.

The way we constructed our formalism 
working from string theory and only 
throwing away though a null set of 
information, it is of course a priori 
expected, that our formalism should be
string theory. In so far there sufficient 
holes in the ``derivation'' of our 
formalism from string theory to be 
equivalent to the latter, that we still 
need the more indirect support from 
rederiving features of string theory 
such as the Veneziano amplitude from our 
model.

Our model is a formulation 
in terms of what we called ``objects'',
and they ``sit'' in circular ``cyclically
ordered chains'', to an open string is 
assigned one such circular chain of 
objects, to a closed string two. The 
``objects'' are supposed to ``sit'' as 
smoothly as they can from quantum 
fluctuations - which put severe 
constraints though, since  {\em the 
odd numbered ``objects'' in cyclically 
ordered chain are not independent 
dynamical variables, but rather given in
terms of (the conjugate variables $\Pi_R^i$
for) the neighboring ``even numbered 
objects'' by equation (\ref{oddfromeven})}.
      
Actually we even stressed that the 
smoothness or continuity condition 
because of the dependence of the odd objects on 
{\em differences} of the conjugate 
momenta of neighboring even ones become 
non-reflection invariant. That is to say
that a cyclically ordered chain being 
smooth would not remain smooth, if one 
puts the objects in the opposite order!      
The crux of the matter is that we have 
a genuine string field theory in the sense
that we construct a state space of Hilbert
vectors describing a whole universe in a 
string theory governed world. Then of 
course there can in the various states of
this Hilbert space exist different numbers
of strings, well this is not hundred percent true, because contrary to other string 
field theories: our Hilbert space is 
described in terms of the ``even objects''
and the number of strings perfectly accurately given once you have a Hilbert space 
state. The ``even objects'' can namely
be associated to strings in slightly 
different ways, so that the number of strings {\em only approximately } can be 
derived from a given state; even there are
no exact eigenstates for the number of 
strings. But in practice we believe the
approximate access to the number of 
strings in our description is sufficient.
But that the number of strings is 
{\em not} cleanly defined feature of a 
state in our Hilbert space, is clear from
the fact that we have scattering even 
scattering that change the number of 
strings, such as if two strings scattered 
and became three, but that nothing 
happen in our formalism under a 
scattering. We just obtained the Veneziano
model scattering amplitude as an {\em 
overlap} of initial and final state just 
corresponding to that nothing happens
in the object formulation. In this sense 
the strings resulting from the 
scattering must have been 
there all the time.   

One may look at our model as {\em solution}
of string theory  in the sense that we 
have ``even object '' description that 
does not even develop with time so that 
the ``even object'' state is more like 
a system of initial data to a solution of
string theory.   

\subsection{Outlook}
We foresee that there must be really very 
much it would be reasonable to do in our
formalism, which is in many ways simpler 
than usual string theory especially than 
usual string field theory. 

Presumably it will be very easy to make
the superstring version; if nothing else
should work one could in principle 
bosonize the fermionic modes and then 
treat the resulting bosons similar to the
way we treated in our model of the bosonic 
modes. 

Of course we should really also properly 
finish getting the Veneziano model 
calculation remaining details. A special
interest might be connected with the 
overall normalization, which we left 
completely out here, since our formalism 
has no obvious candidate for the string 
coupling $g$, so the latter should come 
out from whatever parameters such as our 
cut off parameter $a$ and $\alpha'$ and 
possible vacuum
characteristic, but we did not use 
openly vacuum properties in the 
calculation sketched.

Most interesting might be to use our 
formalism to obtain a better understanding
of the Maldacena conjecture by developing
our formalism for the Ads space and then 
see that the corresponding CFT can also 
be written by our formalism.   
\section*{Acknowledgements}
One of us (HBN) thanks the Niels Bohr Institute for allowing him to stay as emeritus,
and Matias Breskvar for economic support 
to visit the Bled Conference.

The other one(M.N.) acknowledges to 
The Niels Bohr Institute for warmest
hospitality extended to him during his 
stay at NBI. He is supported by the
JSPS Grant in Aid for Scientific 
Research No. 24540293.


\end{document}